\documentclass[preprint,12pt]{elsarticle}

\usepackage{graphicx}

\usepackage{stmaryrd}
\usepackage{amsmath}
\usepackage{amssymb}

\usepackage{todonotes}
\usepackage{tikz}
\usetikzlibrary{arrows,decorations.text,backgrounds,shadows,calc}
\usetikzlibrary{intersections, patterns,automata,arrows,positioning}
\usetikzlibrary{matrix}
\usepackage{varwidth}
\usetikzlibrary{external}
\usepackage{standalone}
\usepackage{tablefootnote}
\usepackage[normalem]{ulem}
\usepackage{xspace}
\usepackage{listings,soul}
\usepackage{subfig}
\usepackage{graphicx}
\usepackage{enumitem}
\usepackage{booktabs}
\usepackage{caption}
\usepackage{tcolorbox}
\usepackage{breakcites}
\usepackage{enumitem}
\usepackage{array}

\usepackage{hyperref}

\usepackage{color}

\usepackage[group-digits=integer, group-minimum-digits=4,
list-final-separator={, and },
free-standing-units, unit-optional-argument, 
detect-weight=true,
per-mode=fraction, 
]{siunitx}
\robustify{\bfseries} 

\usepackage[noend]{algpseudocode}
\usepackage{algorithm}
\algrenewcommand\algorithmiccomment[2][\itshape]{{#1\hfill\(\triangleright\)
		#2}}
\algrenewcommand{\algorithmicrequire}{\textbf{Input:}}
\algrenewcommand{\algorithmicensure}{\textbf{Output:}}
\algrenewcommand{\algorithmicforall}{\textbf{for each}}
\algnewcommand{\IfThenElse}[3]{
	\State \algorithmicif\ #1\ \algorithmicthen\ #2\ \algorithmicelse\ #3}

\usepackage{varwidth}

\usepackage{cleveref}
\usepackage{bold-extra} 
\usepackage{wrapfig}

\crefname{figure}{Fig.}{Fig.}
\Crefname{figure}{Figure}{Figures}
\crefname{definition}{Def.}{Defs.}
\Crefname{definition}{Definition}{Definitions}
\crefname{table}{Tab.}{Tab.}
\Crefname{table}{Table}{Tables}
\crefname{equation}{Eq.}{Eq.}
\Crefname{equation}{Equation}{Equations}
\crefname{algorithm}{Alg.}{Alg.}
\Crefname{algorithm}{Algorithm}{Algorithms}
\crefname{section}{Sec.}{Sec.}
\Crefname{section}{Section}{Sections}

\newcommand{\linefill}{\cleaders\hbox{$\smash{\mkern-2mu\mathord-\mkern-2mu}$}\hfill\vphantom{\lower1pt\hbox{$\rightarrow$}}}  
  
\newcommand{\transi}[2]{\mathrel{\lower1pt\hbox{$\mathrel-_{\vphantom{#2}}\mkern-8mu\stackrel{#1}{\linefill_{\vphantom{#2}}}\mkern-11mu\rightarrow_{#2}$}}}
\newcommand{\trans}[1]{\transi{#1}{{}}}

\newcommand{\cpatrans}[2]{\overset{#2}{\rightsquigarrow}_{#1}}
\newcommand{\cpatransindex}[1]{\cpatrans{#1}{{}}}

\newcommand{\inlineheadingbf}[1]{\medskip\noindent{\bfseries #1.}}
\newcommand{\inlineheadingit}[1]{\medskip\noindent{\it #1.}}

\tikzstyle{ca}=[draw,circle,fill=black!5,rounded corners]  
\tikzstyle{caf}=[draw,double, circle,fill=black!5,rounded corners] 
\tikzstyle{leer} = [rectangle,node distance=.6cm]              

\makeatletter
\pgfdeclareshape{document}{
	\inheritsavedanchors[from=rectangle] 
	\inheritanchorborder[from=rectangle]
	\inheritanchor[from=rectangle]{center}
	\inheritanchor[from=rectangle]{north}
	\inheritanchor[from=rectangle]{south}
	\inheritanchor[from=rectangle]{west}
	\inheritanchor[from=rectangle]{east}
	\backgroundpath{
		\southwest \pgf@xa=\pgf@x \pgf@ya=\pgf@y
		\northeast \pgf@xb=\pgf@x \pgf@yb=\pgf@y
		\pgf@xc=\pgf@xb \advance\pgf@xc by-10pt 
		\pgf@yc=\pgf@yb \advance\pgf@yc by-10pt
		\pgfpathmoveto{\pgfpoint{\pgf@xa}{\pgf@ya}}
		\pgfpathlineto{\pgfpoint{\pgf@xa}{\pgf@yb}}
		\pgfpathlineto{\pgfpoint{\pgf@xc}{\pgf@yb}}
		\pgfpathlineto{\pgfpoint{\pgf@xb}{\pgf@yc}}
		\pgfpathlineto{\pgfpoint{\pgf@xb}{\pgf@ya}}
		\pgfpathclose
		\pgfpathmoveto{\pgfpoint{\pgf@xc}{\pgf@yb}}
		\pgfpathlineto{\pgfpoint{\pgf@xc}{\pgf@yc}}
		\pgfpathlineto{\pgfpoint{\pgf@xb}{\pgf@yc}}
		\pgfpathlineto{\pgfpoint{\pgf@xc}{\pgf@yc}}
	}
}
\makeatother

\definecolor{colorActors}{HTML}{EEEEEE}

\definecolor{colorArtifacts}{HTML}{FFFFFF}  
\definecolor{colorConcepts}{HTML}{EEEEEE}
\definecolor{colorRangedAnalysis}{HTML}{F8961E}
\definecolor{colorPaths}{HTML}{F3722C}
\definecolor{colorPaths2}{HTML}{43AA8B}
\definecolor{colorLower}{HTML}{8CB369}
\definecolor{colorUpper}{HTML}{BC4B51}
\definecolor{colorTask}{HTML}{EEEEEE}
\newcommand{\nameColorLower}{green\xspace}
\newcommand{\nameColorUpper}{red\xspace}

\tikzset{>=latex}
\tikzstyle{ent}=[draw,fill=lGreen, text width=18mm,
text centered,minimum width=20mm, minimum height=3em, rounded corners, drop shadow]
\tikzstyle{inv}=[draw, fill=green!30!white, text width=4em,
text centered,minimum width=4em, minimum height=1em, rounded corners, drop shadow]
\tikzstyle{ent-small}=[draw, fill=lGreen, text width=3.5em,
text centered,minimum width=3em, minimum height=2em, rounded corners, drop shadow]
\tikzstyle{ent-mapper}=[ent-small, fill=lGreen]     
\tikzstyle{ent-large}=[draw, fill=sBlue, text width=8em,
text centered,minimum width=8em, minimum height=4em, rounded corners, drop shadow,
rectangle split, rectangle split draw splits=false, rectangle split parts=2]

\tikzstyle{doc}=[%
draw,
thick,
align=center,
color=black,
fill=sBlue,
font=\footnotesize,
shape=document,
minimum width=10mm,
text width=12mm,
minimum height=8mm,
shape=document,
inner sep=1mm,
]
\tikzstyle{artifact}=[%
draw,
thick,
align=center,
color=black,
fill=yellow,
shape=document,
minimum width=8mm,
minimum height=10mm,
shape=document,
inner sep=1mm,
]

\tikzset{
	verdict/.style={align=center},
	component/.style={draw, align=center, text width=2.2cm, minimum width=2.5cm, minimum height=1cm, },
	tall/.style={},
	actor/.style={component, rectangle, rounded corners=20pt, fill=colorActors},
	concept_small/.style={draw, align=center, text width=1.8cm, minimum width=1.8cm, minimum height=1cm, rectangle, rounded corners=1pt, fill=colorConcepts},
	ra_small/.style={draw, align=center, text width=2.2cm, minimum width=2.2cm, minimum height=1cm, rectangle, rounded corners=1pt, fill=colorRangedAnalysis},
	tool_small/.style={draw, align=center, text width=1.5cm, minimum width=1.5cm, minimum height=1cm, rectangle, rounded corners=1pt, fill=colorTool},
	concept/.style={component, rectangle, rounded corners=1pt, fill=colorConcepts},
	artifact/.style={draw, fill=white, document, rounded corners=1pt,
		minimum width=.7cm, minimum height=0.9cm, inner sep=1pt, align=center, text height=5mm},
	flow/.style={->, thick, >=latex},
	cyc/.style={bend left=30},
	verdict/.style={align=center},
	largetext/.style={font=\large},
	cex/.style={fill=red!30!white},
	paths/.style={fill=yellow!30!white},
	invariant/.style={fill=green!30!white},
	witnessNode/.style={state, fill=white, inner sep=0pt, align=center},
	witnessInvariantNode/.style={witnessNode, rectangle, rounded corners=5pt, inner sep=2pt},
	witnessEdge/.style={draw, ->},
	witnessLabel/.style={font=\ttfamily, align=left},
	witnessLabelOw/.style={font=\itshape},
	initial text=,
}

\let\origthelstnumber\thelstnumber
\makeatletter

\renewcommand\thelstnumber{%
	\ifnum\value{lstnumber}>-1
	\origthelstnumber
	\else
	\ifnum\value{lstnumber}=-1
	$$
	\fi
	\fi
}

\newcommand{\highlightInARG}[2]{\colorbox{#1}{#2}}

\usetikzlibrary{shadows}
\tikzset{
fill lower half/.style={path picture={\fill[#1] (path picture bounding box.south west)
		rectangle (path picture bounding box.east);}},
fill upper half/.style={path picture={\fill[#1] (path picture bounding box.north west)
		rectangle (path picture bounding box.east);}},
fill both halfs/.style 2 args={fill=#2, path picture={\fill[#1] (path picture bounding box.south west)
		rectangle (path picture bounding box.east);}},
reversed diagonal fill/.style 2 args={fill=#2, path picture={
		\fill[#1, sharp corners] (path picture bounding box.north west) |- 
		(path picture bounding box.south east) -- cycle;}}
}

\newcommand{\VnodeDistance}{3mm}

\newcommand{\scaleForFig}{.75}

\newcommand{\CFA}{P\xspace}
\newcommand{\Loc}{Loc\xspace}
\newcommand{\isIf}{B\xspace}

\newcommand{\bexpr}{\mathit{BExpr}\xspace}

\newcommand{\state}{c\xspace}
\newcommand{\mergeOp}{\textsf{merge}}
\newcommand{\stopOp}{\textsf{stop}}

\newcommand{\pathOrder}{\leq}

\newcommand{\noLowerBound}{\pi_{^{_\bot}}}
\newcommand{\noUpperBound}{\pi^{_{\top}}}
\newcommand{\rangeReduction}{\mathbb{R}}
\newcommand{\inducedTau}{\pi_{\tau}}
\newcommand{\inducedTauOne}{\pi_{\tau_1}}
\newcommand{\inducedTauTwo}{\pi_{\tau_2}}
\newcommand{\rangeReductionNoUpper}{\rangeReduction_{[\inducedTauOne, \noUpperBound]}}

\newcommand{\rangeReductionNoLower}{\rangeReduction_{[\noLowerBound,\inducedTauTwo]}}
\newcommand{\rangeReductionUpperLower}{\rangeReduction_{[\inducedTauOne, \inducedTauTwo]}}

\newcommand{\tool}[1]{{\small\textsc{#1}}\xspace}
\newcommand{\cpachecker}{\tool{CPA\-checker}}
\newcommand{\esbmc}{\tool{Esbmc}}
\newcommand{\ultimate}{\tool{UltimateAutomizer}}
\newcommand{\mopsa}{\tool{Mopsa}}
\newcommand{\coveriteam}{\tool{Co\-Veri\-Team}}
\newcommand{\benchexec}{\tool{BenchExec}}

\newcommand{\svcomp}{\tool{SV-Comp}}
\newcommand{\testcomp}{\tool{Test-Comp}}
\newcommand{\svbenchmarks}{\tool{SV-Benchmarks}}

\newcommand{\SymbExec}{\textsc{Se}\xspace}
\newcommand{\BMC}{\textsc{Bmc}\xspace}
\newcommand{\Pred}{\textsc{Pred}\xspace}
\newcommand{\Value}{\textsc{Value}\xspace}
\newcommand{\UA}{\textsc{UAutomizer}\xspace}

\newcommand{\random}{\textsc{Rdm}}
\newcommand{\randomNinetyTen}{\textsc{Rdm9}}

\newcommand{\LBThree}{\textsc{Lb3}\xspace}
\newcommand{\LBTen}{\textsc{Lb10}\xspace}

\newcommand{\WorkStealing}{\textsc{Ws}\xspace}
\newcommand{\Inv}{\mathit{Inv}}

\newcommand{\RASE}{\textsc{Ra-\SymbExec}\xspace}
\newcommand{\RASEPred}{\textsc{Ra-\SymbExec-\Pred}\xspace}
\newcommand{\WSSEPred}{\textsc{Ws-\SymbExec-\Pred}\xspace}
\newcommand{\RAValue}{\textsc{Ra-\Value}\xspace}
\newcommand{\RAValuePred}{\textsc{Ra-\Value-\Pred}\xspace}
\newcommand{\WSValuePred}{\textsc{Ws-\Value-\Pred}\xspace}
\newcommand{\RABMC}{\textsc{Ra-\BMC}\xspace}
\newcommand{\RABMCPred}{\textsc{Ra-\BMC-\Pred}\xspace}
\newcommand{\WSBMCPred}{\textsc{Ws-\BMC-\Pred}\xspace}
\newcommand{\RAPred}{\textsc{Ra-\Pred}\xspace}

\newcommand{\zenodo}{\tool{Zenodo}}
\newcommand{\correctProofSE}{573}
\newcommand{\correctAlarmSE}{912}
\newcommand{\correctOverallSE}{1485}
\newcommand{\incorrectProofSE}{0}
\newcommand{\incorrectAlarmSE}{24}

\newcommand{\correctProofRASE}{567}
\newcommand{\correctAlarmRASE}{948}
\newcommand{\correctOverallRASE}{1515}
\newcommand{\incorrectProofRASE}{0}
\newcommand{\incorrectAlarmRASE}{54}

\newcommand{\additionalVsFirstRASE}{101}
\newcommand{\moreSolvedVsFirstRASE}{30}
\newcommand{\moreSolvedPercentageVsFirstRASE}{2.0}
\newcommand{\correctProofPred}{1281}
\newcommand{\correctAlarmPred}{1209}
\newcommand{\correctOverallPred}{2490}
\newcommand{\incorrectProofPred}{0}
\newcommand{\incorrectAlarmPred}{36}

\newcommand{\correctProofRAPred}{1243}
\newcommand{\correctAlarmRAPred}{1138}
\newcommand{\correctOverallRAPred}{2381}
\newcommand{\incorrectProofRAPred}{0}
\newcommand{\incorrectAlarmRAPred}{48}

\newcommand{\additionalVsFirstRAPred}{62}

\newcommand{\correctProofValue}{827}
\newcommand{\correctAlarmValue}{826}
\newcommand{\correctOverallValue}{1653}
\newcommand{\incorrectProofValue}{0}
\newcommand{\incorrectAlarmValue}{20}

\newcommand{\correctProofRAValue}{829}
\newcommand{\correctAlarmRAValue}{913}
\newcommand{\correctOverallRAValue}{1742}
\newcommand{\incorrectProofRAValue}{0}
\newcommand{\incorrectAlarmRAValue}{48}

\newcommand{\additionalVsFirstRAValue}{189}
\newcommand{\moreSolvedVsFirstRAValue}{89}
\newcommand{\moreSolvedPercentageVsFirstRAValue}{5.4}
\newcommand{\correctProofBMC}{933}
\newcommand{\correctAlarmBMC}{1774}
\newcommand{\correctOverallBMC}{2707}
\newcommand{\incorrectProofBMC}{0}
\newcommand{\incorrectAlarmBMC}{59}

\newcommand{\correctProofRABMC}{923}
\newcommand{\correctAlarmRABMC}{1725}
\newcommand{\correctOverallRABMC}{2648}
\newcommand{\incorrectProofRABMC}{0}
\newcommand{\incorrectAlarmRABMC}{59}

\newcommand{\additionalVsFirstRABMC}{121}

\newcommand{\correctProofRASEPred}{868}
\newcommand{\correctAlarmRASEPred}{1150}
\newcommand{\correctOverallRASEPred}{2018}
\newcommand{\incorrectProofRASEPred}{0}
\newcommand{\incorrectAlarmRASEPred}{37}

\newcommand{\moreSolvedVsFirstRASEPred}{533}

\newcommand{\moreSolvedVsOtherRASEPred}{472}

\newcommand{\additionalVsBothRASEPred}{31}
\newcommand{\correctProofRAValuePred}{874}
\newcommand{\correctAlarmRAValuePred}{1060}
\newcommand{\correctOverallRAValuePred}{1934}
\newcommand{\incorrectProofRAValuePred}{0}
\newcommand{\incorrectAlarmRAValuePred}{52}

\newcommand{\moreSolvedVsFirstRAValuePred}{281}

\newcommand{\moreSolvedVsOtherRAValuePred}{556}

\newcommand{\additionalVsBothRAValuePred}{34}
\newcommand{\correctProofRABMCPred}{1158}
\newcommand{\correctAlarmRABMCPred}{1532}
\newcommand{\correctOverallRABMCPred}{2690}
\newcommand{\incorrectProofRABMCPred}{0}
\newcommand{\incorrectAlarmRABMCPred}{59}

\newcommand{\moreSolvedVsFirstRABMCPred}{17}

\newcommand{\moreSolvedVsOtherRABMCPred}{200}

\newcommand{\additionalVsBothRABMCPred}{28}

\newcommand{\correctProofWSSEPred}{1293}
\newcommand{\correctAlarmWSSEPred}{1335}
\newcommand{\correctOverallWSSEPred}{2628}
\newcommand{\incorrectProofWSSEPred}{0}
\newcommand{\incorrectAlarmWSSEPred}{37}

\newcommand{\moreSolvedVsFirstWSSEPred}{1143}
\newcommand{\moreSolvedPercentageVsFirstWSSEPred}{77.0}

\newcommand{\moreSolvedVsOtherWSSEPred}{138}
\newcommand{\moreSolvedPercentageVsOtherWSSEPred}{5.5}
\newcommand{\additionalVsBothWSSEPred}{29}
\newcommand{\correctProofWSValuePred}{1105}
\newcommand{\correctAlarmWSValuePred}{1236}
\newcommand{\correctOverallWSValuePred}{2341}
\newcommand{\incorrectProofWSValuePred}{0}
\newcommand{\incorrectAlarmWSValuePred}{52}

\newcommand{\moreSolvedVsFirstWSValuePred}{688}

\newcommand{\moreSolvedVsOtherWSValuePred}{149}

\newcommand{\additionalVsBothWSValuePred}{38}
\newcommand{\correctProofWSBMCPred}{1612}
\newcommand{\correctAlarmWSBMCPred}{1617}
\newcommand{\correctOverallWSBMCPred}{3229}
\newcommand{\incorrectProofWSBMCPred}{0}
\newcommand{\incorrectAlarmWSBMCPred}{59}

\newcommand{\moreSolvedVsFirstWSBMCPred}{522}
\newcommand{\moreSolvedPercentageVsFirstWSBMCPred}{19.3}

\newcommand{\moreSolvedVsOtherWSBMCPred}{739}
\newcommand{\moreSolvedPercentageVsOtherWSBMCPred}{29.7}
\newcommand{\additionalVsBothWSBMCPred}{33}

\newcommand{\additionalVsFirstWSSEPredVsRA}{634}

\newcommand{\additionalVsFirstWSValuePredVsRA}{416}

\newcommand{\additionalVsFirstWSBMCPredVsRA}{564}

\newcommand{\numberTasksValidatedByBoth}{469}
\newcommand{\numberTasksSolvedRAOnReducedDataset}{463}
\newcommand{\numberTasksValidatedRAOnReducedDataset}{460}
\newcommand{\numberTasksNotValidatedRAOnReducedDataset}{3}
\newcommand{\percentagValidatedWitnessesRAOnReducedDataset}{99.4}

\journal{Science of Computer Programming}

\begin{document}

\begin{frontmatter}

\title{Parallel Program Analysis on Path Ranges}
\author[uol]{Jan Haltermann\fnref{fn1}}
\author[lmu]{Marie-Christine Jakobs}
\author[uol]{Cedric Richter}
\author[uol]{Heike Wehrheim\fnref{fn1}}

\fntext[fn1]{Partially supported by the German Research Foundation (DFG) — \href{https://gepris.dfg.de/gepris/projekt/418257054?context=projekt&task=showDetail&id=418257054&}{WE2290/13-1 (Coop)}.}

\affiliation[uol]{organization={Department of Computing Science, Carl von Ossietzky Universitat Oldenburg},
            addressline={Ammerlaender Heerstra\ss{}e 114-118}, 
            city={Oldenburg},
            postcode={26129}, 
            country={Germany}}
\affiliation[lmu]{organization={Department of Computer Science, LMU Munich},
	addressline={Oettingenstraße 67}, 
	city={80538},
	postcode={Munich}, 
	country={Germany}}

\begin{abstract}
Symbolic execution is a software verification technique symbolically running programs and thereby checking for bugs.  
Ranged symbolic execution  
performs symbolic execution on program parts, so-called {\em path ranges}, in parallel. 
Due to the parallelism, verification is accelerated and hence scales to larger programs. 

In this paper, we discuss a generalization of ranged symbolic execution to arbitrary program analyses.
More specifically, we present a verification approach that splits programs into path ranges and then runs arbitrary analyses on the ranges in parallel. 
Our approach in particular allows to run {\em different} analyses on different program parts. 
We have implemented this generalization on top of the tool \textsc{CPAchecker} and evaluated it on programs from the SV-COMP benchmark. Our evaluation shows that verification can benefit from the parallelization of the verification task, but also needs a form of work stealing (between analyses) to become efficient.

\end{abstract}


\end{frontmatter}



\section{Introduction}
In recent years, automatic software verification has become a mature technique reaching levels of industrial software.
This progress is achieved by verification tools employing a bunch of different techniques for analysis, like predicate abstraction, bounded model checking, k-induction, property-directed reachability, or automata-based methods. 
As individual tools still have their specific strength and weaknesses, recent research in software verification has studied {\em cooperative verification}~\cite{DBLP:conf/isola/Beyer20}.

In cooperative verification, tools together work on verification tasks, typically by verifying different parts of the software.
This principle has already been implemented in various forms~\cite{DBLP:conf/icse/BeyerHLW22,DBLP:conf/icse/BeyerJLW18,DBLP:conf/fm/ChristakisMW12,CoVEGI}, in particular also as cooperations of testing and verification~\cite{CoVeriTest,CheckNCrash,CzechConditionalModelChecking,AbstractionConcolicTesting}. 
Cooperative verification often takes the form of a sequential (or cyclic) combination in which tools run in some sequential order, passing information from one to the next tool. In such a sequential combination, the first tool might try to verify the entire program, and if it does not succeed, passes information about the unverified part to the next tool, which continues verification. 
An example of this is conditional model checking~\cite{ConditionalModelChecking,CzechConditionalModelChecking,DBLP:conf/icse/BeyerJLW18} in which the first tool passes information about the unverified part in the form of a so-called condition. 
A condition has an automaton-like structure and can encode (parts of) programs. 
In such approaches, the division of work (i.e., the split of the program) is not apriori computed, but instead implicitly determined by the first tool. 

{\em Parallel} combinations (in which different tools operate in parallel)   in the majority of cases employ {\em portfolio} approaches, simply running different tools on the same task in parallel. 
In a portfolio approach, there is no need for splitting programs:
every tool gets exactly the same program, and typically the first tool completing verification determines the outcome of the verification task. 
It is even possible to run several instances of the same tool in parallel, just with different configurations.  
The reason for mainly employing portfolio approaches for parallel combinations is often the lack of appropriate splitting techniques, which could divide the program into parts on which analyses could be run in parallel.

The proposal of Siddiqui and Khurshid~\cite{Sym-Range} for {\em ranged symbolic execution} provided exactly one such splitting technique. 
They defined so-called {\em path ranges} as parts of the program on which several symbolic execution instances can be run in parallel. 
A path range describes a set of program paths defined by two inputs to the program, where the path $\pi_1$ triggered by the first input is the lower bound and the path $\pi_2$ for the second input is the upper bound of the range. 
All paths in between, i.e.,~paths $\pi$ such that $\pi_1 \! \pathOrder\! \pi\! \pathOrder\! \pi_2$   (based on some ordering $\mathord{\pathOrder}$ on paths), make up a range. 
Then, several symbolic execution engines can be run in parallel on different ranges, searching for bugs in path ranges.  
Siddiqui and Khurshid experimentally confirmed that symbolic execution can benefit from this splitting into path ranges and 
this helps to scale the technique to larger programs.

In this paper, we generalize ranged symbolic execution to arbitrary analyses.
Our generalization concerns the usage of different analyses. 
Instead of its specialization to one analysis, namely symbolic execution, we present ranged program analysis in which arbitrary analyses (in particular also different ones) can be run on the different ranges of a program in parallel. 
In this, we employ a form of {\em work stealing} in which one analysis can -- once done with its own range -- take over ranges assigned to other analyses to better balance workloads. 
We have furthermore extended the approach of \cite{Sym-Range} by a new splitting technique.
Instead of only choosing ranges randomly, we propose a splitting technique along loop bounds of arbitrary, user-configurable numbers $n$. 
A loop bound $n$ splits program paths into ranges only entering the loop at most $n$ times and ranges entering loops for more than $n$ times\footnote{Such splits can also be performed on intervals on loop bounds, thereby generating more than two path ranges.}.

Finally, our approach incorporates the  {\em joining} of results obtained on ranges. 
In ranged symbolic execution, bugs found in ranges are simply collected and reported as bugs for the entire program. 
We keep this form of bug joining and complement it by the joining of correctness results. 
Specifically, we provide a way of joining {\em correctness witnesses}~\cite{CorrectnessWitnesses15} certifying the absence of bugs in path ranges 
into correct witnesses for the entire program.

We have implemented our technique of ranged analyses within the configurable software verification tool \textsc{CPAchecker}~\cite{CPAchecker}. \textsc{CPAchecker} already provides a number of analyses, all defined as configurable program analyses (CPAs). To steer the analyses to particular ranges, we implemented a specific range reduction analysis CPA and then employed the built-in feature of analysis composition to 
combine it with different analyses. 
Our evaluation on SV-COMP 2023 benchmarks~\cite{svbenchmarks23} shows in particular, that
path-based analyses like symbolic execution and value analysis benefit most from being used within ranged program analysis.
Moreover, when using the novel technique of work stealing within ranged program analysis, the combination of two different analyses solves more tasks than the standalone analyses.
Finally, we experimentally confirmed the soundness of the technique for joining correctness witnesses.

\smallskip
\noindent 
This paper builds on and extends our previous publication~\cite{RangedAnalysisFASE}. 
The extension concerns the novel integration of work stealing into ranged analysis and the joining of witnesses. 
We have furthermore significantly extended our evaluation to see which improvements the novel concepts bring and to validate witness joining. Our experiments build on the result of the evaluation described in \cite{RangedAnalysisFASE}; we in particular 
do not repeat the experiments for different splitting strategies as the previous experiments have already shown which splitting strategy is the -- on average -- best one. 
	

\section{Background}

	The task in software verification is to decide, whether a given program satisfies a certain property.
	In this work, we aim to check if a C~program can violate an assertion. 
	More precisely, the analyses aim to show the unreachability of an assertion failure. 
To be able to formally analyze programs, we start by introducing some background concerning the syntax and semantics of programs and proceed by defining path ranges as well as the configurable program analyses implemented in \cpachecker.

\subsection{Program Syntax and Semantics}
In this paper, we consider simple, imperative programs with a deterministic control-flow with integer variables only (taken from some overall set of variables \(\mathcal{V}\))\footnote{Our implementation supports C~programs.}.
A program is formally modeled by a \emph{control-flow auto\-maton}~(CFA) \(\CFA=(L,\ell_0, G)\), where \(L\subseteq \Loc\) is a subset of the program locations~\(\Loc\) (the program counter values), \(\ell_0\in L\) is the initial location, and control-flow edges \(G\subseteq L\times Ops\times L\) describe the execution of operations in locations leading to next locations. 
Therein, the set of operations~\(Ops\) comprises assume statements (Boolean expressions over variables~\(\mathcal{V}\), denoted by $\bexpr$), assignments, {\tt{skip}} (empty) operations, and assertions. 
We expect that CFAs originate from program code and, thus, control-flow may only branch at assume operations, i.e., CFAs~\(\CFA=(L,\ell_0, G)\) are deterministic in the following sense: 
For all \((\ell,op',\ell'), (\ell,op'',\ell'')\in G\) either \(op'=op''\wedge \ell'=\ell''\) or  \(op', op''\) {are assume operations} and \(op'\equiv \neg(op'')\).
We assume that there exists an {\em indicator function} $\isIf_\CFA: G \rightarrow \{T, F, N\}$ that reports the branch direction, either N(one), T(rue), or F(alse).
This indicator function assigns \(N\) to all edges without assume operations and for any two assume operations  \((\ell,op',\ell'), (\ell,op'',\ell'')\in G\) with \(op'\neq op''\) it guarantees \(\isIf_\CFA((\ell,op',\ell'))\, \cup \, \isIf_\CFA((\ell,op'',\ell''))=\{T,F\}\).
Since CFAs are typically derived from programs and assume operations correspond to the two evaluations of conditions of e.g.,~if or while statements, the assume operation representing the true evaluation of the condition is typically assigned \(T\). We will later need this indicator function for defining path orderings.

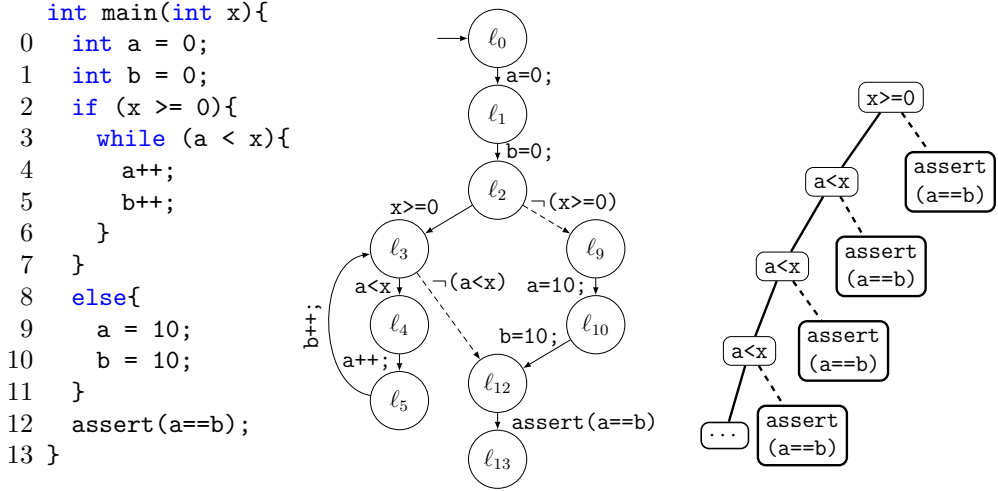
\begin{figure}[t]
	\begin{minipage}[b]{0.25\textwidth}
\begin{lstlisting}
int main(int x){
  int a = 0;
  int b = 0;
  if (x >= 0){ 
    while (a < x){ 
      a++;
      b++;
    }
  }
  else{
    a = 10;
    b = 10;
  }
  assert(a==b); 
}
\end{lstlisting}
	\end{minipage}
	\begin{minipage}[b]{0.33\textwidth}
		\label{fig:example-cfa}
		\scalebox{\scaleForFig}{
			\newcommand{\HnodeDistanceLarge}{10mm}
			\begin{tikzpicture}[node distance=\VnodeDistance]
				\node[witnessNode, initial,initial where=left] (2) {$\ell_0$};
				\node[witnessNode, below = \VnodeDistance of 2] (3) {$\ell_1$};
				\node[witnessNode, below = \VnodeDistance of 3] (4) {$\ell_2$};
				\node[witnessNode, below left= \VnodeDistance and \HnodeDistanceLarge of 4] (5) {$\ell_3$};
				\node[witnessNode, below=  \VnodeDistance of 5] (6) {$\ell_{4}$};
				\node[witnessNode, below = \VnodeDistance of 6] (7) {$\ell_5$};
				\node[witnessNode, below right= \VnodeDistance and \HnodeDistanceLarge of 4] (11) {$\ell_{9}$};
				\node[witnessNode, below = \VnodeDistance of 11] (12) {$\ell_{10}$};

				\node[witnessNode, below left= \VnodeDistance and \HnodeDistanceLarge of 12] (15) {$\ell_{12}$};
				\node[witnessNode, below = \VnodeDistance of 15] (16) {$\ell_{13}$};

				\path
				(2) edge[witnessEdge] node[right, witnessLabel] {a=0;} (3)
				(3) edge[witnessEdge] node[right, witnessLabel] {b=0;} (4)
				(4) edge[witnessEdge] node[left, witnessLabel, pos=0.05, xshift=-4mm] {x>=0} (5)
				(4) edge[witnessEdge, densely dashed] node[above, witnessLabel, pos=0.5, xshift=5mm] {$\neg$(x>=0)} (11)
				(5) edge[witnessEdge, densely dashed] node[right, witnessLabel, pos=0.1] {$\neg$(a<x)} (15)
				(5) edge[witnessEdge, ] node[left, witnessLabel, ] {a<x} (6)
				(6) edge[witnessEdge] node[left, witnessLabel] {a++;} (7)
				(7) edge[witnessEdge, bend left=80] node[above, rotate=90, witnessLabel, pos=0.5,yshift=0mm, xshift=0mm] {b++;} (5)
				(11) edge[witnessEdge] node[left, witnessLabel] {a=10;} (12)
				(12) edge[witnessEdge] node[above, witnessLabel,xshift=-3mm] {b=10;} (15)
				(15) edge[witnessEdge] node[right, witnessLabel, xshift=1mm] {assert(a==b)} (16)
				;
			\end{tikzpicture}
		}
	\end{minipage}  \hfill
	\begin{minipage}{0.34\textwidth}
		\vspace*{-5.5cm}
		\scalebox{\scaleForFig}{
		\begin{tikzpicture}[level/.style={sibling distance = 3.5cm/#1,
				level distance = 1.5cm,},
			ir/.style = {shape=rectangle, rounded corners,font=\ttfamily,
				draw, align=center, color=black,  },
			oor/.style = {draw=gray, color=gray },
			oorNode/.style= {shape=rectangle, rounded corners,
				draw=gray, align=center, color=gray,font=\ttfamily}]]
			\node[ir] (start) {x>=0}
			child{ node[ir, xshift=7mm](end){a<x}  [black, very thick]
				child{ node[ir, thin]{a<x} [black, very thick]
					child{ node[ir, thin]{a<x} [black, very thick]
						child{ node[ir, thick]{$\cdots$} [ir,very thick]}
						child{ node[ir,  black, xshift=0.5cm] {assert\\(a==b)} [dashed, black, very thick]}
					}
					child{ node[ir,  black, xshift=0.5cm] {assert\\(a==b)} [dashed, black, very thick]}
				}
				child{ node[ir, black,, very thick ]{assert\\(a==b)} [dashed, black,]}
			}
			child{ node[ir, black, xshift=-7mm, very thick]{assert\\(a==b)} [dashed,black, very thick]
			};
			
		\end{tikzpicture}
		}
	\end{minipage}
	\caption{Example program, its CFA, and a shortened execution tree}
	\label{fig:exCFA}
	 \vspace{-5mm}
\end{figure}
\Cref{fig:exCFA} (left) shows our example program, which either iteratively increments two variables or directly assigns the constant value \num{10} to them.
For each condition of an if or while statement, the CFA (shown in the middle) contains one assume edge for each evaluation of the condition, depicted as solid edges labeled by the condition for entering the if branch or loop when the condition evaluates to true and dashed edges labeled by the negated condition for entering the else branch or leaving the loop when the condition evaluates to false. 
All other statements are represented by a single edge. The assertion {\tt{assert(a==b)}} at the end of the program is the 
property to be checked for the program, and here we see that the program is correct. 

Next, we give the semantics of programs. 
A program \emph{state} is a pair $(\ell,c)$ of a program location~\(\ell\in L\) and a data state~\(c\) from the set~\(C\) of data states that assign to each variable~\(\mathtt{v}\in\mathcal{V}\) a value of the variable's domain.
Program \emph{execution paths} \(\pi=(\ell_0,c_0) \trans{g_1}(\ell_1, c_1) \trans{g_2} \ldots \trans{g_n} (\ell_n, c_n)\) are sequences of states and edges
such that (1)~they start at the beginning of the program and (2) only perform valid execution steps that (a)~adhere to the control-flow, i.e., \(\forall 1\leq i\leq n: g_i=(\ell_{i-1},\cdot, \ell_i)\in G\), and (b)~properly describe the effect of the operations, i.e., \(\forall 1\leq i\leq n: c_i=sp_{op_i}(c_{i-1})\), where the strongest postcondition \(sp_{op_i}: C \rightharpoonup C\) is a partial function modeling the effect of operation~\(op_i \in Ops\) on data states.
Execution paths are also called \emph{feasible} paths, and paths that fulfill properties (1) and (2a) but violate property (2b) are called \emph{infeasible} paths.
The set of all feasible execution paths of a program $\CFA$ is denoted by \(paths(\CFA)\). 

\subsection{Path Ordering, Execution Trees, and Ranges}
Ranged symbolic execution runs on so-called path ranges, which are sets of ``consecutive'' paths.
To specify these sets, we first define an {\em ordering} on (feasible) execution paths. 
Given two program paths \(\pi=(\ell_0,c_0) \trans{g_1}(\ell_1, c_1) \trans{g_2} \ldots \trans{g_n} (\ell_n, c_n)\in paths(\CFA)\) and \(\pi'=(\ell'_0,c'_0) \trans{g'_1}(\ell'_1, c'_1) \trans{g'_2} \ldots \trans{g'_m} (\ell'_m, c'_m)\in paths(\CFA)\), we define their order~\(\pathOrder\) based on their control-flow edges. 
More specifically, edges with assume operations representing a true evaluation of a condition are smaller than the edges representing the corresponding false evaluation of that condition.
Following this idea, \(\pi\pathOrder\pi'\) if
\(\exists \ 0\leq k\leq n: \forall\ 1\leq i\leq k: $ $g_i\!=\!g'_i \wedge \big((n=k \wedge m\geq n) \vee (m>k\wedge n>k\) \(\wedge  \isIf_\CFA(g_{k+1})=T \wedge  \isIf_\CFA(g'_{k+1})\!=\!F)\big)\).
Note that the path order $\pathOrder$ is a total function, as each feasible execution path of a CFA \CFA starts in $\ell_0$.

An \emph{execution tree} is a tree containing all feasible execution paths of a program with the previously defined ordering, where nodes are labeled with the assume operations.
We depict a shortened version of the execution tree for our example program in \cref{fig:exCFA}.

Based on the above ordering, we now specify ranges, which describe sets of ``consecutive'' program execution paths analyzed by a ranged analysis and which are characterized by a left and right path that limits the range.
Hence, a \emph{range} \([\pi, \pi']\) is the set \(\{\pi_r\in paths(\CFA)\mid \pi\pathOrder \pi_r\pathOrder \pi'\}\)\footnote{In~\cite{Sym-Range}, the range is formalized as \([\pi, \pi')\) but their implementation works on \([\pi, \pi']\).}.
To easily describe ranges that are not bound on the left or right, we use the special paths \(\noLowerBound, \noUpperBound \notin paths(\CFA)\) which are smaller and greater than every path, i.e., \(\forall \pi\in paths(\CFA): (\pi\pathOrder \noUpperBound)\)  \(\wedge~ (\noUpperBound\not\pathOrder\pi) \wedge (\noLowerBound\pathOrder\pi) \wedge (\pi\not\pathOrder  \noLowerBound)\).
Consequently, \([\noLowerBound,\noUpperBound]=paths(\CFA)\).

As the program is assumed to be deterministic except for the input, a \emph{test case} $\tau$, $\tau: \mathcal{V} \rightarrow \mathbb{Z}$,
which maps each input variable to a concrete value,
describes exactly a single path $\pi$\footnote{More concretely, test input~\(\tau\) describes a single maximal path and all its prefixes.}.
We say that $\tau$ {\em induces}
$\pi$ and write this path as $\pi_\tau$.
Consequently, we can define a range by two induced paths, i.e., as
\([\inducedTauOne, \inducedTauTwo]\) for two test cases $\tau_1$ and $\tau_2$.
For the example program from \cref{fig:exCFA}, two example test cases are $ {\tau_1} = \{x\mapsto 2\}$
and ${\tau_2} = \{x\mapsto 0\}$.
Two such induced paths are 
\begin{align*}
	\inducedTauOne =  (\ell_0,\{x\!\mapsto\!2\}) \! & \trans{a=0}          & (\ell_1,~   & \{x\!\mapsto\!2,\!a\!\mapsto\!0\})               \\
	                                                & \trans{b=0}          & (\ell_2,~   & \{x\!\mapsto\!2,\!a\!\mapsto\!0,b\!\mapsto\!0\}) \\
	                                                & \trans{x>=0}         & (\ell_{3},~ & \{x\!\mapsto\!2,\!a\!\mapsto\!0,b\!\mapsto\!0\}) \\
	                                                & \trans{a<x}          & (\ell_{4},~ & \{x\!\mapsto\!2,\!a\!\mapsto\!0,b\!\mapsto\!0\}) \\
	                                                & \trans{a++}          & (\ell_{5},~ & \{x\!\mapsto\!2,\!a\!\mapsto\!1,b\!\mapsto\!0\}) \\
	                                                & \trans{b++}          & (\ell_{3},~ & \{x\!\mapsto\!2,\!a\!\mapsto\!1,b\!\mapsto\!1\}) \\
	                                                & \trans{a<x}          & (\ell_{4},~ & \{x\!\mapsto\!2,\!a\!\mapsto\!1,b\!\mapsto\!1\}) \\
	                                                & \trans{a++}          & (\ell_{5},~ & \{x\!\mapsto\!2,\!a\!\mapsto\!2,b\!\mapsto\!1\}) \\
	                                                & \trans{b++}          & (\ell_{3},~ & \{x\!\mapsto\!2,\!a\!\mapsto\!2,b\!\mapsto\!2\}) \\
	                                                & \trans{\neg(a<x)}    & (\ell_{15}, & \{x\!\mapsto\!2,\!a\!\mapsto\!2,b\!\mapsto\!2\}) \\
	                                                & \trans{assert(a==b)} & (\ell_{16}, & \{x\!\mapsto\!2,\!a\!\mapsto\!2,b\!\mapsto\!2\})
\end{align*}
and
\begin{align*}
	\inducedTauTwo = (\ell_2,\{x\!\mapsto\!0\}) \, 	  & \trans{a=0}          & (\ell_1,~   & \{x\!\mapsto\!0,\!a\!\mapsto\!0\})                  \\
	                                                  & \trans{b=0}          & (\ell_3,~   & \{x\!\mapsto\!0,\!a\!\mapsto\!0,b\!\mapsto\!0\})      \\
	                                                  & \trans{x>=0}         & (\ell_{3},~ & \{x\!\mapsto\!0,\!a\!\mapsto\!0,b\!\mapsto\!0\}) \\
	                                                  & \trans{\neg(a<x)}    & (\ell_{12}, & \{x\!\mapsto\!0,\!a\!\mapsto\!0,b\!\mapsto\!0\}) \\
	                                                  & \trans{assert(a==b)} & (\ell_{13}, & \{x\!\mapsto\!0,\!a\!\mapsto\!0,b\!\mapsto\!0\})  .
\end{align*}

\subsection{Configurable Program Analysis}\label{sec:cpa-background}
We will realize our ranged analysis using the \emph{configurable program analysis}~(CPA) framework~\cite{CPA}. This framework allows the definition of customized, abstract-inter\-pretation based analyses, i.e., it allows a definition of an abstract domain and transfer functions for it as well as configurations steering the exploration of the abstract state space. 
For the latter, one defines when and how to combine abstract states and when to stop exploration.
Formally, a CPA~\(\mathbb{A}=(D,\rightsquigarrow, \mergeOp, \stopOp)\) consists of
\begin{itemize}
	\item the \emph{abstract domain}~\(D=(\Loc\times C, (E,\top,\sqsubseteq,\sqcup),\llbracket\cdot\rrbracket)\), which is composed of a set~\(\Loc\times C\) of program states, a join semi-lattice on the abstract states~\(E\) as well as a concretization function $\llbracket\cdot\rrbracket$ fulfilling  
	      \[\llbracket\top\rrbracket=\Loc\times C \mathrm{~and~} \forall e, e'\in E: \llbracket e\rrbracket\cup\llbracket e'\rrbracket\subseteq\llbracket e\sqcup e'\rrbracket\]
	\item the \emph{transfer relation}~\(\mathord{\rightsquigarrow} \subseteq E\times {G} \times E\) defining the abstract semantics that safely overapproximates the program semantics, i.e.,
	      \begin{align*} &\forall e\in E,  \ g\in \Loc\times Ops\times\Loc:                                                                                                                                   \\                              
															&\{(\ell',c')\mid \exists \mathrm{~valid~execution~step~} (\ell,c)\trans{g}(\ell',c'): (\ell,c)\in\llbracket e\rrbracket\}\subseteq \bigcup_{(e,g,e')\in\rightsquigarrow}\llbracket e'\rrbracket
	      \end{align*}
	\item the \emph{merge operator}~\(\mergeOp: E\times E\rightarrow E\) used to combine information satisfying 
	      \[\forall e, e'\in E: e'\sqsubseteq\mergeOp(e,e')\]
	\item the \emph{termination check}~\(\stopOp: E\times 2^E\rightarrow \mathbb{B}\) deciding whether the exploration of an abstract state can be omitted and fulfilling 
	      \[\forall e\in E, E_\mathrm{sub}\subseteq E: \stopOp(e,E_\mathrm{sub})\implies \llbracket e\rrbracket\subseteq\bigcup_{e'\in E_\mathrm{sub}}\llbracket e'\rrbracket\]
\end{itemize}
To run the configured analysis, one executes a meta-reachability analysis, the so-called CPA algorithm, configured by the CPA. 
To this end, an initial value $e_{init} \in E$ which the analysis will start with has to be given.
For details on the CPA algorithm, we refer the reader to \cite{CPA}.

\inlineheadingbf{Value Analysis}
To restrict our ranged analysis to a particular program range, we track the induced paths defining the left and right border of the range using the abstract domain and transfer relation of a CPA~\(\mathbb{V}\) for \emph{value analysis}~\cite{CPA+} (also known as constant propagation or explicit analysis).
An abstract state~\(v\) of the value analysis ignores program locations and maps each variable to either a concrete value of its domain or \(\top\), which represents any value.
The partial order~\(\sqsubseteq_\mathbb{V}\) and the join operator \(\sqcup_\mathbb{V}\) are defined variable-wise while ensuring that \(v\sqsubseteq _\mathbb{V} v' \Leftrightarrow \forall \mathtt{v}\in\mathcal{V}: v(\mathtt{v})=v'(\mathtt{v}) \vee v'(\mathtt{v})=\top\)\footnote{Consequently, \(\forall \mathtt{v}\in\mathcal{V}:\top_\mathbb{V}(\mathtt{v})=\top\).} and \((v\sqcup_\mathbb{V} v')(\mathtt{v})=v(\mathtt{v})\) if \(v(\mathtt{v})=v'(\mathtt{v})\) and otherwise \((v\sqcup_\mathbb{V} v')(\mathtt{v})=\top\).
The concretization of abstract state~\(v\) contains all concrete states that agree on the concrete variable values, i.e., \(\llbracket v\rrbracket_\mathbb{V}:=\left\{(\ell,\state)\in \Loc\times C\mid \forall \mathtt{v}\in\mathcal{V}: v(\mathtt{v})=\top\vee v(\mathtt{v})=\state(\mathtt{v})\right\}\).
If the values for all relevant variables are known, the transfer relation $ \cpatransindex{\mathbb{V}}$ will behave like the program semantics.
Otherwise, it may overapproximate the executability of a CFA edge and may assign value~\(\top\) if a concrete value cannot be determined.

\inlineheadingbf{Composite CPA}
To easily build ranged analysis instances for various program analyses, we modularize our ranged analysis into a {\em range reduction} (see \cref{sec:range-reduction}) and a program analysis.
Technically, we will compose a ranged analysis from different CPAs using the concept of a \emph{composite CPA}~\cite{CPA}.
Here, we just define the composition of two CPAs. 
The composition of more than two CPAs works analogously or can be achieved by recursively composing two (composite) CPAs.
A composite CPA~\(\mathbb{A}_\times=(D_\times,\cpatransindex{\times}, \mergeOp_\times, \stopOp_\times)\)
of CPA~\(\mathbb{A}_1=((\Loc\times C, (E_1,\top_1,\sqsubseteq_1,\) \(\sqcup_1),\llbracket\cdot\rrbracket_1),\cpatransindex{1}, \mergeOp_1, \stopOp_1)\) and
CPA~\(\mathbb{A}_2=((\Loc\times C, (E_2,\top_2,\sqsubseteq_2,\sqcup_2),\llbracket\cdot\rrbracket_2), \cpatransindex{2},\) \(\mergeOp_2, \stopOp_2)\)
considers the product domain \(D_\times\!=\!(\Loc\!\times \!C,(E_1\!\times\! E_2, (\top_1, \) \(\top_2),\) 
\(\sqsubseteq_\times,\sqcup_\times), \llbracket\cdot\rrbracket_\times)\)
that defines the operators elementwise, i.e., for pairs of abstract states we define \((e_1,e_2)\!\sqsubseteq_\times \!(e'_1, e'_2)\) if \(e_1\!\sqsubseteq_1 \!e'_1\)
and \(e_2\!\sqsubseteq_2\! e'_2\), \((e_1,e_1)\sqcup_\times (e'_1, e'_2)\!=\!(e_1\!\sqcup_1\! e'_1, e_2\!\sqcup\! e'_2)\),
and \(\llbracket (e_1, e_2)\rrbracket\!=\!\llbracket e_1\rrbracket_1\!\cap\!\llbracket e_2\rrbracket_2\).
The trans\-fer relation may be the product transfer relation or may strengthen the product transfer relation using knowledge about the other abstract successor.
Similarly, we may define \(\mergeOp_\times\) element\-wise, i.e., \(\mergeOp_\times((e_1,e_2), (e'_1, e'_2))=(\mergeOp_1(e_1, e'_1), \mergeOp_2(e_2, e'_2))\) or define a customized, more precise merge operator. 
While a conjunction of the individual termination checks is typically not sound,  in practice most composite analyses use the termination check \(\stopOp_\times(e,E_\mathrm{sub}):=\stopOp_\mathrm{sep}(e,E_\mathrm{sub}):=\exists e'\in E_\mathrm{sub}: e\sqsubseteq_\times e'\), which can be automatically derived from the partial order.
\section{Ranged Program Analysis} \label{sec:combined-range-analysis}

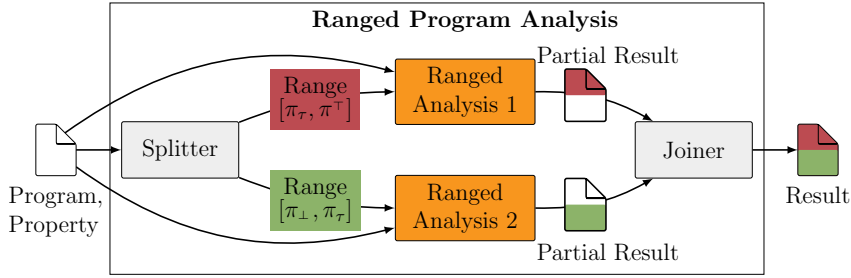
\begin{figure}[t]
	\centering
	\newcommand{\hdiff}{20mm}
	\newcommand{\hdiffWide}{30mm}
	\newcommand{\vdiff}{3mm}
	\scalebox{0.75}{
		\begin{tikzpicture}
			\pgfdeclarelayer{bg}
			\pgfsetlayers{bg,main}
			\node[ concept_small] (split) {Splitter};
			\node[right= \hdiffWide of split.east, draw=none, minimum width=2cm] (ra2){};
			\node[ ra_small, above=\vdiff of ra2] (ra1) {Ranged Analysis 1};
			\node[ ra_small, below=\vdiff of ra2] (ra4) {Ranged Analysis 2};
			\node[ concept_small, right = \hdiff of ra2.east] (aggregation) {Joiner};
			\node[artifact, fill=white, left=8mm of split.west] (prog) {};
			\node[below = 0.5mm of prog.south, anchor=north,align=left] (prog-label) {Program,\\Property};	
			\node[artifact,   fill both halfs={colorLower}{colorUpper},right=8mm of aggregation.east](res) {};
			\node[below = 0.5mm of res.south, anchor=north] (res-label) {Result};			
			\node[above = 1cm of ra1, anchor=north] (orchestration) {\textbf{Ranged Program Analysis}};	
			
			\path (prog) edge[flow]  (split);			
			\path (split) edge[flow,bend left=10,] node[verdict, fill=colorUpper] {Range\\[-1mm] [$\inducedTau, \noUpperBound$]}   (ra1.west);
			\path (split) edge[flow,bend right=10,]node[verdict, fill=colorLower] {Range\\[-1mm][$\noLowerBound, \inducedTau$] } (ra4.west);		
			\path (ra1.east) edge[flow,bend left=15,pos=0.4] node[artifact,fill upper half=colorUpper,label={[xshift=0.4cm, yshift=-0.05cm]Partial Result}] {} (aggregation);
			
			\path (ra4.east) edge[flow,bend right=15,pos=0.4] node[artifact,fill lower half=colorLower,label={[xshift=0.4cm, yshift=-1.6cm]Partial Result}] {} (aggregation);
			\path (aggregation) edge[flow]  (res);
			\begin{pgfonlayer}{bg}
				
				\draw[black, draw, fill=white] ($(split.north west) + (-.2, 2.1)$) rectangle ($(aggregation.south east) + (0.2,-1.7)$) ;
				\path ($(prog.east) + (-.2,+.3) $) edge[flow,bend left=25,]  ($(ra1.west) + (0,+.35) $);
				\path ($(prog.east) + (0,-.3) $) edge[flow,bend right=25, ] ($(ra4.west) + (0,-.35) $);
			\end{pgfonlayer}

 \end{tikzpicture}}

	\caption{Conceptual overview of ranged program analysis for two ranged analyses
		\label{fig:overview-ranged-analysis}}	
		\vspace*{-5mm}
\end{figure}

In this section, we introduce {\em ranged program analysis} for generalizing the concept of ranged symbolic execution to arbitrary program analyses. 
Our goal is to split the set of program paths in a given program into multiple disjoint ranges, each of which can be analyzed by an independent program analysis. 
For the analysis of a given range, we introduce the concept of {\em ranged analysis}. 
A ranged analysis is a program analysis whose exploration is limited to a specific range. We realize ranged analysis as a composition of an arbitrary analysis and a {\em range reduction}, both defined as a configurable program analysis (CPA). 
Then, to analyze the complete program with a ranged program analysis, we (1) {\em split} the program into ranges, (2) run several ranged analyses in parallel, and (3) {\em aggregate} the analysis results.  
A conceptual overview of the ranged program analysis of two ranges is shown in \cref{fig:overview-ranged-analysis}.
Next, we formalize ranged analysis in \cref{sec:ranged-analysis}.
Then, we define in \cref{sec:range-reduction} the range reduction, which is used to restrict an arbitrary program analysis to the paths within a range.
Thereafter, we explain how to handle underspecified test cases in \cref{sec:underspec-testcases} and introduce the splitter \LBThree for generating ranges in \cref{sec:gen-ranges}.

\subsection{Ranged Analysis}\label{sec:ranged-analysis}
We show that any program analysis can be transformed into a ranged analysis by composing
it with our novel range reduction. The range reduction can be defined as a configurable program analysis (CPA)
that explores all paths included in a range $[\inducedTauOne,\inducedTauTwo]$ and aborts on all other paths. By composing
the range reduction with another analysis, the composed ana\-lysis only explores paths that are feasible in both analyses, thus, excluding paths that are out of range. 

We decompose the range reduction for $[\inducedTauOne,\inducedTauTwo]$ into a composition of two specialized ranged reductions $\rangeReductionNoUpper$ and $\rangeReductionNoLower$, which decide whether a path is in the range $[\inducedTauOne, \noUpperBound]$ and 
$[\noLowerBound,\inducedTauTwo]$, respectively. 
Since $[\inducedTauOne, \inducedTauTwo] = [\inducedTauOne, \noUpperBound] \cap [\noLowerBound, \inducedTauTwo]$
and the composition stops the exploration of a path if one analysis returns $\bot$ (i.e., no states to explore),
the composite analysis $\rangeReductionUpperLower =\rangeReductionNoLower \times \rangeReductionNoUpper$ 
only explores paths that are included in both ranges (which are 
exactly the paths in $[\inducedTauOne, \inducedTauTwo]$). 

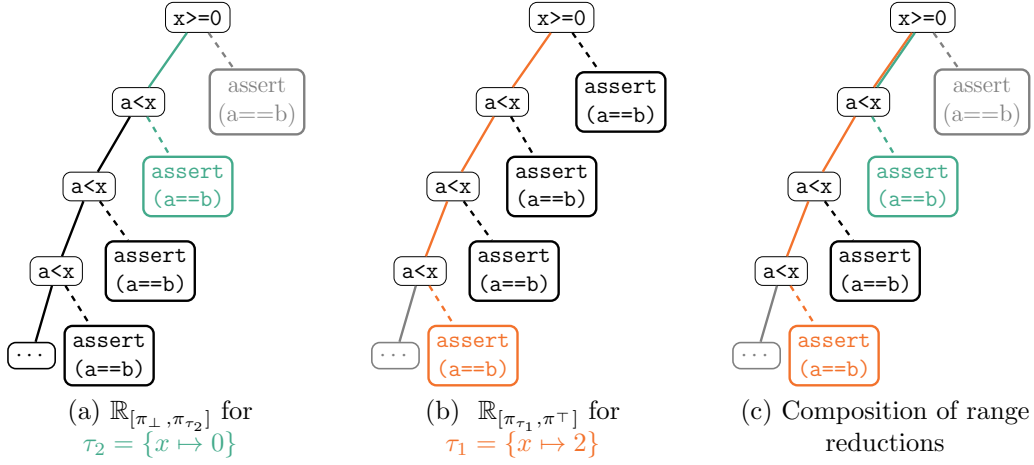
\begin{figure}[t]
    \captionsetup[subfigure]{justification=centering}
	\subfloat[$\rangeReductionNoLower$ for \\ {\color{colorPaths2} ${\tau_2} = \{x\mapsto 0\}$}\label{fig:upperBounded}]{\scalebox{\scaleForFig}{
\begin{tikzpicture}[level/.style={sibling distance = 3.5cm/#1,
				level distance = 1.5cm,},
		ir/.style = {shape=rectangle, rounded corners,font=\ttfamily,
				draw, align=center, color=black, },
	oor/.style = {draw=gray, color=gray },
oorNode/.style= {shape=rectangle, rounded corners,
	draw=gray, align=center, color=gray,}]]
	\node[ir] {x>=0}
	child{ node[ir, xshift=7mm]{a<x}  [black, very thick, colorPaths2]
			child{ node[ir, thin]{a<x} [black, very thick]
				child{ node[ir, thin]{a<x} [black, very thick]
					child{ node[ir, thick]{$\cdots$} [ir,very thick]}
					child{ node[ir,  black, xshift=0.5cm] {assert\\(a==b)} [dashed, black, very thick]}
				}
					child{ node[ir,  black, xshift=0.5cm] {assert\\(a==b)} [dashed, black, very thick]}
				}
			child{ node[ir, colorPaths2,, very thick ]{assert\\(a==b)} [dashed, colorPaths2,]}
		}
	child{ node[oorNode,  xshift=-7mm, very thick]{assert\\(a==b)} [dashed,oor, very thick]
		};
\end{tikzpicture}}}\hfill
	\subfloat[ $\rangeReductionNoUpper$ for \\ {\color{colorPaths} $ {\tau_1} = \{x\mapsto 2\}$}\label{fig:lowerBounded}]{\scalebox{\scaleForFig}{
\begin{tikzpicture}[level/.style={sibling distance = 3.5cm/#1,
		level distance = 1.5cm,},
	ir/.style = {shape=rectangle, rounded corners,font=\ttfamily,
		draw, align=center, color=black, },
	oor/.style = {draw=gray, color=gray },
oorNode/.style= {shape=rectangle, rounded corners,
	draw=gray, align=center, color=gray,}]]
	\node[ir] (start) {x>=0}
	child{ node[ir, xshift=7mm](end){a<x}  [colorPaths, very thick]
		child{ node[ir, thin]{a<x} [colorPaths, very thick]
			child{ node[ir, thin]{a<x} [colorPaths, very thick]
				child{ node[oorNode, thick]{$\cdots$} [oor,very thick]}
				child{ node[ir,  colorPaths, xshift=0.5cm] {assert\\(a==b)} [dashed, colorPaths, very thick]}
			}
			child{ node[ir,  black, xshift=0.5cm] {assert\\(a==b)} [dashed, black, very thick]}
		}
		child{ node[ir, black, very thick ]{assert\\(a==b)} [dashed, black, very thick,]}
	}
	child{ node[ir, black, xshift=-7mm, very thick]{assert\\(a==b)} [dashed,black, very thick]
	};

\end{tikzpicture}}}\hfill 
	\subfloat[Composition of range\\ reductions\label{fig:exampleCompositeRanged}]{\scalebox{\scaleForFig}{
\begin{tikzpicture}[level/.style={sibling distance = 3.5cm/#1,
		level distance = 1.5cm,},
	ir/.style = {shape=rectangle, rounded corners,font=\ttfamily,
		draw, align=center, color=black, },
	oor/.style = {draw=gray, color=gray },
	oorNode/.style= {shape=rectangle, rounded corners,
		draw=gray, align=center, color=gray,}]]
	\node[ir] (start) {x>=0}
	child{ node[ir, xshift=7mm](end){a<x}  [colorPaths, very thick]
		child{ node[ir, thin]{a<x} [colorPaths, very thick]
			child{ node[ir, thin]{a<x} [colorPaths, very thick]
				child{ node[oorNode, thick]{$\cdots$} [oor,very thick]}
				child{ node[ir,  colorPaths, xshift=0.5cm] {assert\\(a==b)} [dashed, colorPaths, very thick]}
			}
			child{ node[ir,  black, xshift=0.5cm] {assert\\(a==b)} [dashed, black, very thick]}
		}
		child{ node[ir, colorPaths2,, very thick ]{assert\\(a==b)} [dashed, colorPaths2,]}
	}
	child{ node[oorNode, , xshift=-7mm, very thick]{assert\\(a==b)} [dashed,oorNode, very thick]
	};
	\path ($(start.south) + (-.17,+0) $) edge[,gray, densely dotted, very thin ]($(end.north) + (0.212,0) $);
	\path ($(start.south) + (-.14,+0) $) edge[,colorPaths2, very thick]  ($(end.north) + (0.242,0) $);
\end{tikzpicture}}}
	\vspace{-2mm}
	\caption{Application of range reduction on the running example of \cref{fig:exCFA}} \label{fig:exampleRanged}
\end{figure} 
\Cref{fig:exampleRanged} depicts the application of range reduction to the example from \cref{fig:exCFA}, where the range reduction $\rangeReductionNoLower$ is depicted in \cref{fig:upperBounded} and 
$\rangeReductionNoUpper$ in \cref{fig:lowerBounded} and
the composition of both range reductions in \cref{fig:exampleCompositeRanged}.
Finally, to transform any arbitrary program analysis $\mathbb{A}$ into a ranged analysis, we construct the following composition for a given range $[\inducedTauOne, \inducedTauTwo]$:
\begin{equation*}
	\rangeReductionNoUpper \times \rangeReductionNoLower \times \mathbb{A}
\end{equation*}
For $\rangeReductionUpperLower$, we use \(\stopOp_{sep}\) and define \(\mergeOp_\times\) component-wise for the individual merge operators as explained in \cref{sec:cpa-background}.

\subsection{Range Reduction as CPA}\label{sec:range-reduction}
Next, we define the range reduction $\rangeReductionNoUpper$ ($\rangeReductionNoLower$, respectively) as a CPA, which tracks whether a state is reached via a path in $[\inducedTauOne, \noUpperBound]$ ($[\noLowerBound, \inducedTauTwo]$).

\inlineheadingbf{Initialization} 
To define the CPAs for $\rangeReductionNoUpper$ and $\rangeReductionNoLower$, 
we reuse components of the value analysis $\mathbb{V} 
$ (as described in~\cref{sec:cpa-background}). 
A value analysis explores at least all feasible paths of a program by tracking the values for program variables. 
If the program behavior is fully determined (i.e.,~all (input) variables are set to constants), then only one feasible, maximal path exists, which is explored by the value analysis.
We exploit this behavior by initializing the analysis based on our test case $\tau$ (being a lower or upper bound of a range): \vspace{-2mm} 
\begin{equation*}
	e_{init} = \left\{
\begin{array}{ll}
v(x) = \tau(x) & \text{ if } x \in dom(\tau), x \in \mathcal{V} \\
v(x) = \top & \, \text{otherwise} \\
\end{array}
\right.
\end{equation*}

\vspace{-1mm}
In this case, all variables which are typically undetermined\footnote{We assume that randomness is controlled only through an input and hence the program is deterministic.} and dependent on the program input have now a determined value, defined through the test case.
As the behavior of the program under the test case $\tau$ is now fully determined, the value analysis only explores a single path $\pi_\tau$,
which corresponds to the execution trace of the program given the test case.
Now, as we are interested in all paths defined in a range and not only a single path, we adapt the value analysis as follows:

\inlineheadingbf{Lower Bound CPA}
For the CPA range reduction $\rangeReductionNoUpper$,
we borrow all components of the value analysis except 
for the transfer relation $\cpatransindex{\tau_1}$. 
The transfer relation $\cpatransindex{\tau_1}$ is defined as follows:\vspace{-2mm}
\[(v, g, v') \in \mathord{\cpatrans{\tau_1}{}} \text{ iff }\begin{cases}
	 v=\top \wedge v' = \top, \text{or} \\
	  v \neq \top \wedge \ v' = \top \wedge  \isIf_\CFA(g) = F \wedge  (v,g,\bot) \in  \mathord{\cpatrans{\mathbb{V}}{}}, \text{or} \\
	 v\neq \top \wedge \big(  v' \neq \bot \vee \isIf_\CFA(g) \neq F \big) \wedge  (v,g,v') \in \mathord{\cpatrans{\mathbb{V}}{}}
\end{cases}\]

 \vspace{-1mm}
\noindent Note that $\top$ represents the value analysis state where no information on variables is stored
and $\bot$ represents an unreachable state in the value analysis, which stops the exploration of the path.
Hence, the second case ensures that $\rangeReductionNoUpper$ also visits the false-branch of a condition when the path induced by $\tau_1$ follows the true-branch. 
Note that in case that $\cpatrans{\mathbb{V}}{}$ computes $\bot$ as a successor state for an assumption $g$ with $\isIf_\CFA(g) = T$, the exploration of the path is stopped, as $\inducedTauOne$ follows the false-branch (contained in the third case).

\inlineheadingbf{Upper Bound CPA}
For the CPA range reduction $\rangeReductionNoLower$, we again borrow all components of the value analysis except  for the 
transfer relation $\cpatransindex{\tau_2}$.
The transfer relation $\cpatransindex{\tau_2}$ is defined as follows: \vspace{-2mm}
\[(v,g,  v') \in \mathord{\cpatrans{\tau_2}{}} \text{ iff } \begin{cases}
	  v=\top \wedge v' = \top \\
	  v \neq  \top \wedge  v' = \top \wedge \isIf_\CFA(g) = T \wedge  (v,g,\bot) \in  \mathord{\cpatrans{\mathbb{V}}{}} \\
 v\neq \top \wedge \big(  v' \neq \bot \vee \isIf_\CFA(g) \neq T \big) \wedge  (v,g,v') \in \mathord{\cpatrans{\mathbb{V}}{}}
\end{cases}\]

\vspace{-1mm}
\noindent
The second condition now ensures that $\rangeReductionNoLower$ also visits the true-branch
of a condition when $\inducedTauTwo$ follows the false-branch.

\subsection{Handling Underspecified Test Cases}\label{sec:underspec-testcases}
So far, we have assumed that test cases are fully specified, i.e.,~contain values for all input variables, and the behavior of the program is deterministic such that executing a test case $\tau$ follows a single (maximal) execution path $\pi_\tau$.
However, in practice, we observe that test cases can be underspecified such that a test 
case $\tau$ does not provide concrete values for all input variables.
We denote by $P_\tau$ the {\em set} of all paths that are then induced by $\tau$. 
In this case, we define:\vspace{-2mm}
\[\vspace*{-2mm}
[\noLowerBound, P_\tau] = \{\pi \mid \forall \pi^\prime \in P_\tau: \pi \pathOrder\pi^\prime\} = \{\pi \mid \pi \pathOrder \text{min}(P_\tau)\} 
\]

\vspace{-2mm}
\noindent and 
\begin{equation*}
[P_\tau, \noUpperBound] = \{\pi \mid \exists \pi^\prime \in P_\tau:  \pi^\prime \leq \pi\} = \{\pi \mid \text{min}(P_\tau) \leq \pi\}
\end{equation*}
Interestingly enough, by defining $\pi_\tau = \text{min}(P_\tau)$ for an underspecified test case~$\tau$ we can handle the range as if $\tau$ would be fully specified.

\subsection{Splitting}\label{sec:gen-ranges}
To be able to apply ranged program analysis, generating ranges, the splitting of programs into parts that can be analyzed in parallel, is a crucial part.
For splitting, a range can either be defined by two program paths or two test cases that need to be computed.
We propose and evaluate four different splitting in~\cite{RangedAnalysisFASE}.
The evaluation shows that the strategy called \LBThree generates ranges leading to the best performance of ranged program analysis.
Thus, we will only explain \LBThree in detail and only briefly summarize the other three.

\inlineheadingbf{The Splitter \LBThree}
The goal of this splitter is to generate a range that contains only three loop iterations.
More precisely, the splitter \LBThree computes the left-most path in the program that contains exactly three unrollings of the loop.
If the program contains nested loops, each nested loop is unrolled three times in each iteration of the outer loop.
To generate the test case for the computed path, we (1) build its path formula using the strongest post-condition operator~\cite{strongestPost}, 
(2) use an SMT-solver to check the formula for satisfiability and (3) in case of an answer ``SAT'', use the evaluation of the input variables provided by the SMT-solver in the path formula as one test case. 
In case the path formula is unsatisfiable, we shorten the path by
iteratively removing the last statement from it, until we get a satisfying path formula. 
A test case~$\tau$ generated by \LBThree defines the two ranges $[\noLowerBound,\pi_\tau]$ and $[\pi_\tau,\noUpperBound]$.  
For loop-free programs, we cannot bind the number of loop unrollings.
Hence, \LBThree fails and we generate a single range $[\noLowerBound, \noUpperBound]$.
In our running example in  \cref{fig:exCFA}, \LBThree~generates $\tau=\{x\mapsto 3\}$. 

\inlineheadingbf{Other splitting strategies}
Beneath \LBThree we also introduce the splitting strategy \LBTen, which is also based on loop unrollings.
In contrast to \LBThree, \LBTen computes the left-most path that contains exactly ten loop unrollings.
The other two strategies select paths randomly by traversing the CFA and deciding at each assume edge to either follow the true- or the false-branch.
\random~selects both branches with a probability of \num{50}\%, whereas \randomNinetyTen~selects the true-branch with a \num{90}\% probability.
As execution trees of programs with loops are often not balanced but rather grow to the left, \randomNinetyTen~likely generates longer paths.
For a more detailed description, we refer the reader to~\cite{RangedAnalysisFASE}.

\section{Joining Witnesses}\label{sec:witness-join}
At the end of each ranged program analysis, we need to aggregate the partial results of all ranged analyses (see Fig.~\ref{fig:overview-ranged-analysis}) into a joint result since each of them only analyzes a subset of the program behavior.
We explained in~\cite{RangedAnalysisFASE} how to compute a joint verdict.
In addition to the verdict, analyses often generate witnesses~\cite{CorrectnessWitnesses15,ViolationWitness15,WitnessJournal}, which provide explanations for the respective verdicts.
However, each ranged analysis can only produce a witness that explains the verdict of its own analysis and, thus, can only give an explanation for the subset of program behavior it analyzed.
Therefore, we discuss how to join the witnesses generated by the ranged analyses into a joint witness for the complete program.

First, let us look at the verdict ``false'', i.e., the ranged program analysis reports that the program violates the analyzed property.
For the verdict ``false'', it is sufficient that at least one of the ranged analyses reports ``false''. 
This analysis also produces a violation witness~\cite{ViolationWitness15,WitnessJournal} that describes a counter\-example path, which witnesses the violation.
Furthermore, its violation witness is sufficient to explain the aggregated verdict ``false'' of the ranged program analysis.
If more than one ranged program analysis computes a violation witness, reporting all paths violating the property may be beneficial for users and developers, as one path might be shorter or simpler than the other.
We execute the ranged analyses in parallel and currently abort the other ranged analyses as soon as one ranged analysis detects a counterexample.
Waiting for the other ranged analyses to (potentially) compute another counterexample requires defining and measuring such an additional timeout while ensuring that the ranged program analysis reports the known verdict ``false'' before it times out.
To keep the concept of ranged program analysis simple, we decided against joining violation witnesses and simply using one of the violation witnesses provided by a ranged analysis with the verdict ``false''.
More concretely, we use the first violation witness being generated. 
Adding a configurable handling of multiple violation witnesses, e.g., by joining them or selecting one heuristically, is planned as future work.

Next, let us consider the verdict ``true'', i.e., the ranged program analysis proves the property.
In this case, all ranged analyses successfully proved that their subset of the program behavior fulfills the property. 
To explain their results, they each provide a correctness witness~\cite{CorrectnessWitnesses15,WitnessJournal}.
Formally, a \emph{correctness witness} for a CFA \(P=(L,\ell_0, G)\) is a (protocol) automaton \(\mathcal{A}=(Q,2^G\times\Phi,\delta, q_0, \Inv)\), which uses the set~\(\Phi\) of all quantifier-free predicate-logic formulae over the program variables to describe sets of data states.
Similar to most automata, a correctness witness consists of a set of states~\(Q\) including initial state \(q_0\in Q\), an alphabet \(2^G\times\Phi\) whose symbols define which operations (i.e., CFA edges) and data states are considered by an automaton transition, and a transition relation \(\delta\subseteq Q\times(2^G\times\Phi)\times Q\).
In addition, it contains a total function \(\Inv:Q\rightarrow\Phi\) that assigns to each state of the automaton an invariant describing an expected set of data states.
Furthermore, a correctness witness must fulfill the following completeness property, which guarantees that the correctness witness does not exclude any program behavior.
\[\forall q\in Q, g\in G: (\bigvee \{\psi\mid \exists q\xrightarrow{(S,\psi)}q'\in\delta: g\in S\})\equiv true\]

\begin{figure}[t]

	\scalebox{0.7}{
		\begin{tikzpicture}[node distance=2.5cm,>=latex]
			\node (s) {};
			\node (q0) [below of =s, node distance=2cm, draw, rounded corners, text width=4em, align=center,font=\footnotesize] {$\ell_0$\\ \highlightInARG{colorLower}{true}\\ \highlightInARG{colorUpper}{true}\\ true};
			\node (q1) [below of = q0, draw, rounded corners, text width=4em, align=center,font=\footnotesize] {$\ell_1$\\ \highlightInARG{colorLower}{true}\\ \highlightInARG{colorUpper}{true}\\ true};
			\node (q2) [below of = q1, draw, rounded corners, text width=6em, align=center,font=\footnotesize] {$\ell_2$\\ \highlightInARG{colorLower}{true}\\ \highlightInARG{colorUpper}{$a=0\wedge b=0$}\\true};
			\node (i)[below of =q2] {};
			\node (q3) [left of = i, node distance=3cm, draw, rounded corners, text width=8em, align=center,font=\footnotesize] {$\ell_3$\\ \highlightInARG{colorLower}{$b=a$}\\ \highlightInARG{colorUpper}{$a=0\wedge b=0$}\\ $b=a \vee a=0\wedge b=0$};
			\node (q4) [right of = i, node distance=3cm, draw, rounded corners, text width=4em, align=center,font=\footnotesize] {$\ell_{9}$\\ \highlightInARG{colorLower}{false}\\ \highlightInARG{colorUpper}{true}\\ true};
			\node (q5) [below of=q3, draw, rounded corners, text width=4em, align=center,font=\footnotesize] {$\ell_4$\\ \highlightInARG{colorLower}{true}\\ \highlightInARG{colorUpper}{false}\\ true};
			\node (q6) [below of=q5, draw, rounded corners, text width=4em, align=center,font=\footnotesize] {$\ell_5$\\ \highlightInARG{colorLower}{true}\\ \highlightInARG{colorUpper}{false}\\ true};
			\node (q7) [below of=q4, draw, rounded corners, text width=4em, align=center,font=\footnotesize] {$\ell_{10}$\\ \highlightInARG{colorLower}{false}\\ \highlightInARG{colorUpper}{true}\\ true};
			\node (q8) [below of=q7, draw, rounded corners, text width=12em, align=center,font=\footnotesize] {$\ell_{12}$\\ \highlightInARG{colorLower}{true}\\ \highlightInARG{colorUpper}{$a=0\wedge b=0\vee a=10\wedge b=10$}\\true};
			\node (q9) [below of=q8, draw, rounded corners, text width=4em, align=center,font=\footnotesize] {$\ell_{13}$\\ \highlightInARG{colorLower}{true}\\ \highlightInARG{colorUpper}{true}\\ true};
			
			\draw[->] (s) -- (q0);
			\draw[->] (q0) to node[left,font=\footnotesize] {(\{\((\ell_0,\texttt{a=0;},\ell_1)\)\}, true)} (q1);	
			\draw[->] (q1) to node[left,font=\footnotesize] {(\{\((\ell_1,\texttt{b=0;},\ell_2)\)\}, true)} (q2);	
			\draw[->] (q2) to node[left,font=\footnotesize] {(\{\((\ell_2,\texttt{x>=0},\ell_3)\)\}, true)} (q3);	
			\draw[->] (q2) to node[right,font=\footnotesize] {(\{\((\ell_2,\texttt{$\neg$(x>=0)},\ell_{9})\)\}, true)} (q4);	
			\draw[->] (q3) to node[left,font=\footnotesize] {(\{\((\ell_3,\texttt{a<x},\ell_4)\)\}, true)} (q5);	
			\draw[->] (q3.east) to node[above,rotate=-50,font=\footnotesize, pos=0.4] {(\{\((\ell_3,\texttt{$\neg$(a<x)},\ell_{12})\)\}, true)} (q8);	
			\draw[->] (q5) to node[left,font=\footnotesize] {(\{\((\ell_4,\texttt{a++;},\ell_5)\)\}, true)} (q6);	
			\draw (q6.east) edge[bend right, ->] node[below,font=\footnotesize, rotate=90] {(\{\((\ell_5,\texttt{b++;},\ell_3)\)\}, true)} (q3);	
			\draw[->] (q4) to node[right,font=\footnotesize] {(\{\((\ell_{9},\texttt{a=10;},\ell_{10})\)\}, true)} (q7);	
			\draw[->] (q7) to node[right,font=\footnotesize] {(\{\((\ell_{10},\texttt{b=10;},\ell_{12})\)\}, true)} (q8);
			\draw[->] (q8) to node[right,font=\footnotesize] {(\{\((\ell_{12},\texttt{assert(a==b);},\ell_{13})\)\}, true)} (q9);	
			
			\draw[->] (q0) edge[loop left] node[left,font=\footnotesize] {(o/w, true)} (q0);
			\draw[->] (q1) edge[loop left] node[left,font=\footnotesize] {(o/w, true)} (q1);
			\draw[->] (q2) edge[loop left] node[left,font=\footnotesize] {(o/w, true)} (q2);
			\draw[->] (q3) edge[loop left] node[left,font=\footnotesize] {(o/w, true)} (q3);
			\draw[->] (q4) edge[loop right] node[right,font=\footnotesize] {(o/w, true)} (q4);
			\draw[->] (q5) edge[loop left] node[left,font=\footnotesize] {(o/w, true)} (q5);
			\draw[->] (q6) edge[loop left] node[left,font=\footnotesize] {(o/w, true)} (q6);
			\draw[->] (q7) edge[loop right] node[right,font=\footnotesize] {(o/w, true)} (q7);
			\draw[->] (q8) edge[loop right] node[right,font=\footnotesize] {(o/w, true)} (q8);
			\draw[->] (q9) edge[loop right] node[right,font=\footnotesize] {(G, true)} (q9);
		\end{tikzpicture}
	}
	\caption{For given test \(\tau_2=\{x\mapsto0\}\) and our example program from Fig.~\ref{fig:exCFA}, correctness witnesses for ranged analysis on  \highlightInARG{colorLower}{[$\noLowerBound, \inducedTauTwo$]} produced by predicate abstraction (\nameColorLower highlighted invariants), correctness witness for ranged analysis on \highlightInARG{colorUpper}{[$\inducedTauTwo, \noUpperBound$]} produced by value analysis (\nameColorUpper highlighted invariants) and correctness witness generated by Alg.~\ref{alg:joinWit} (black invariants) when given these two witnesses and our example program. 
		\label{fig:witnesses}}
\end{figure}

Figure~\ref{fig:witnesses} shows three correctness witnesses for our example program.
For this example, their graph structure is identical and reflects the CFA structure.
However, the witnesses differ in their invariants.
The invariants are the second part of the node labels and are shown in a different color for each witness. 
Note that we use o/w (short for otherwise) to summarize all control-flow edges that do not occur in another outgoing edge of the respective state.
For example, for the initial state o/w stands for \(G\setminus\{(\ell_0, a=0;,\ell_1)\}\).
The witnesses with the \highlightInARG{colorLower}{\nameColorLower} and \highlightInARG{colorUpper}{\nameColorUpper} invariants, respectively, both provide an explanation for the correctness of the subset of the program behavior.
The witness with \nameColorLower invariants might be generated by a ranged analysis based on predicate abstraction~\cite{ABE} considering range [$\noLowerBound, \inducedTauTwo$] resulting from test \(\tau_2=\{x\mapsto0\}\).
The witness with the \nameColorUpper invariants might be produced by a ranged analysis based on value analysis~\cite{ValueAnalysis} considering the opposite range [$\inducedTauTwo, \noUpperBound$].
We observe that both witnesses use the invariant ``false'' for locations that the respective analysis never explored because no paths in their range contain these locations.
Furthermore, note that the analyses use block encoding and, therefore, may only provide invariants for particular locations (either loop heads or locations with multiple outgoing edges).
Locations, which are intermediate locations in the block encoding, are assigned the invariant true.

Each of the two witnesses generated by a ranged analysis provides an explanation for the correctness of the subset of the program behavior analyzed by the respective ranged analysis but likely does not provide sufficient information for behavior analyzed by another ranged analyses.
In the following, we explain how to combine correctness witnesses produced by ranged analyses during ranged program analysis into a joint correctness witness that explains the correctness of the complete program.
The black invariants belong to such a joint witness.
Next, we explain how to compute these joint witnesses.

Algorithm~\ref{alg:joinWit} describes the combination for two correctness witnesses\footnote{The algorithm can easily be extended to combine more than two witnesses. Alternatively, one can for example iteratively combine multiple witnesses by first joining the first two witnesses with Alg.~\ref{alg:joinWit} and then successively using \cref{alg:joinWit} to combine the result of the first n-1 witness combinations with the n-th witness.}.
Its main part (lines 1--9) performs a parallel exploration of the program and the two witnesses.
\begin{algorithm}[t]
\caption{Joining two correctness witnesses}\label{alg:joinWit}
\begin{algorithmic}[1]
\Require  CFA \(P=(L,\ell_0, G)\)
\Require Witnesses \(\mathcal{A}=(Q,2^G\times\Phi,\delta, q_0, \Inv)\),
\Statex\hspace{6em} \(\mathcal{A}'=(Q',2^G\times\Phi,\delta', q'_0, \Inv')\)
\Ensure Joint correctness witness
\State $\mathtt{waitlist}=N=\{(\ell_0,q_0,q'_0)\}$; 
\While{($\texttt{waitlist} \neq\emptyset$)}
  \State pop $(\ell, q, q')$ from \texttt{waitlist}
  \For{each $(\ell, op, \ell_s)\in G$}
	  \For{each $(q, (S,\varphi), q_s)\in \delta$ with \((\ell, op, \ell_s)\in S\)}
		  \For{each $(q', (S', \varphi')', q'_s)\in \delta'$ with \((\ell, op, \ell_s)\in S'\)}
	        \If{$\varphi\wedge\varphi'\not\equiv false\wedge(\ell_s,q_s, q'_s)\notin N$}
		        \State $N = N \cup \{\ell_s,q_s, q'_s\}$; 
					 \State $\texttt{waitlist} = \texttt{waitlist} \cup \{\ell_s,q_s, q'_s\}$;
		      \EndIf
		  \EndFor
		\EndFor
	\EndFor
\EndWhile
\State $\Inv_\mathrm{join} =  \{(\ell, inv) \mid \ell\in L,\hspace{0.5em} inv=false \bigvee_{(\ell, q, q')\in N} (\Inv(q)\vee \Inv(q'))\}$;
\State $\delta_\mathrm{join} = \{(\ell,(\{(\ell, op, \ell')\}, true),\ell')\mid (\ell, op, \ell')\in G\}$
\Statex \hspace{3em}$\cup\{(\ell, (\{(\ell_p, op, \ell_s)\in G\mid \ell_p\neq \ell\},true),\ell)\mid \ell\in L\}$;
\State \Return ($L, 2^G\times\Phi,  \delta_\mathrm{join}, \ell_0, \Inv_\mathrm{join}$);
\end{algorithmic}
\end{algorithm}
While a parallel exploration of the two witnesses would have been sufficient to compute a joint witness, we include the program into the exploration because it allows us to focus on explanations for program behavior, to construct (compact) joint witnesses that align with the program structure, which we assume are better for validation, and to simplify the witness construction.
The parallel exploration starts in the initial location and initial automata states and explores the successors of each explored node once.
Since we only need to provide explanations for program behavior, i.e., program paths, successor exploration focuses on successors that adhere to the program's control-flow.
To this end for each leaving control-flow edge~\(g\) in the program, we inspect all combinations of outgoing transitions of the witnesses that consider edge~\(g\) and that can be taken jointly, i.e., their enabling conditions do not exclude each other.
The resulting successor combinations are then considered further during exploration.
After the parallel exploration, the joint witness is computed.
To avoid that the joint witness forces a validator to unroll the program, which may make it inefficient, \cref{alg:joinWit} computes a rather natural representation of the joint correctness witness, which aligns with the program structure. 
Therefore, line~10 computes one joint invariant per program location.
For each node detected during the parallel exploration, we combine the invariant information from both witnesses.
We choose to combine the information via disjunction.
Thus, we may lose information in case both witnesses consider the full behavior when computing their invariant information.
However, disjunction ensures that we remain sound in case different behaviors are considered.
Based on the combination, the joint invariant for a program location~\(\ell\) is derived from the disjunction of all nodes explored that refer to that location~\(\ell\).
We use a disjunction because different nodes likely represent different behaviors.
Note that we include ``false'' in the disjunction to account for syntactically unreachable locations.
After computing the invariants, line~11 determines the transitions of the joint witness.
We add transitions to encode the program's control-flow.
Due to the format of a transition, we cannot add the control-flow edge directly, but add a transition that accepts the control-flow edge and that is always enabled. 
To make the witness complete, we add self-loops to each state\footnote{More precisely, we add self-loops to each program location} of the joint witness that cover all remaining edges, which do not adhere to the control-flow.
Lastly, line~12 returns the joint witness, which reuses the program's locations and initial location as states and initial states, using the same alphabet as the witnesses and the previously computed transitions and joint invariants.

We implemented \cref{alg:joinWit} for joining of witnesses presented above in the software analysis framework \cpachecker~\cite{CPAchecker}.
For the parallel exploration, we compose a CPA tracking the control-flow with one CPA per witness that tracks the transitions of that witness.
To allow us to combine the witness information of the explored states, we wrap the above composition of CPAs into another CPA (called WrapperCPA), whose transfer relation forwards to the composition. 
We then use \cpachecker's meta-reachability analysis instantiated with the above WrapperCPA to perform the parallel exploration.
Thereafter, we apply \cpachecker's witness generation on the result of the parallel exploration, which relies on the WrapperCPA to combine the witness information and computes the witness similar to \cref{alg:joinWit}.

Finally, let us briefly discuss which invariants a ranged analysis should provide for locations it never explored.
Basically, there are two options: ``true'' or ``false''.
In contrast to ``false'', ``true'' is always a correct invariant with respect to the complete program.
However, correctness witnesses of ranged analyses that only analyze a subset of the program behavior already do not guarantee correct invariants for the complete program for locations they visited during their analysis, especially if they do not analyze all paths to that location.
They only encode information on behavior that they have seen.
Thus, ``false'' would better fit to the information encoded at other seen locations.
In addition, providing ``true'' would result in losing all correctness witness information provided by a different ranged analysis due to the disjunction.
Hence, from the perspective of witness joining ``false'' is preferred.
Since correctness witnesses were not designed for incomplete analysis and do not provide any requirements on how to handle unreached code, it depends on the verifier how it handles locations not analyzed.
Hence, the quality and usefulness of a joint correctness witness depends on how the individual verifiers in the ranged program analysis handle non-analyzed locations and, therefore, may differ amongst verifiers.

\section{Work Stealing}\label{sec:work-stealing}

When employing ranged program analysis with different analyses for the ranges, one has to decide which analysis should work on which range.
As all program analyses have different strengths and weaknesses, the assignment of ranges can affect the performance of the ranged program analysis.
To exemplify this effect, let us revisit the running example from \cref{fig:example-cfa} and assume, that the splitter generates the test case $\tau_2 = \{x \mapsto 0\}$.
Then, the first range $[\noLowerBound, \inducedTauTwo]$ contains all paths within the if-branch and the second range $[\inducedTauTwo, \noUpperBound]$ the else-branch and the path within the if-branch containing no loop unrolling ($\ell_0,\ell_1,\ell_2,\ell_3,\ell_{12},\ell_{13}$).
Now, we employ a ranged analysis using value analysis for the first and a predicate abstraction for the second range.
The predicate abstraction quickly proves the assigned program range safe and terminates.
In contrast, the value analysis has to analyze the program with \num{2147483647} (INT\_MAX) different values for $x$, namely all positive integers from the \texttt{int} domain, as it does not apply any abstraction.
These computations will most likely exceed the given time limitations, resulting in a timeout for the complete ranged program analysis, as only one part of the program has been proven safe.
In contrast to value analysis, predicate abstraction can also prove the first range safe, as it abstracts the loop iterations using a predicate like $a=b$.
Thus, changing the assignment of the program ana\-lyses to the ranges affects the performance, as the value analysis can easily compute a proof for the second range containing the else branch.

\begin{figure}[t]
	\centering
	\newcommand{\hdiff}{12mm}
	\newcommand{\hdiffWide}{25mm}
	\newcommand{\hdiffWider}{40mm}
	\newcommand{\vdiff}{15mm}
	\scalebox{0.7}{
		\begin{tikzpicture}
			\pgfdeclarelayer{bg}
			\pgfsetlayers{bg,main}
			\node[ concept_small] (split) {Splitter};
			\node[right= \hdiffWide of split.east, draw=none, minimum width=2cm] (ra2){};
			\node[ ra_small, above=\vdiff of ra2] (ra1) {Ranged Analysis 1};
			\node[ ra_small, right=\hdiffWider of ra1] (ra1new) {Ranged Analysis 1};
			\node[ ra_small, below=\vdiff of ra2] (ra4) {Ranged Analysis 2};
			\node[ concept_small, right = \hdiff of ra2.east] (aggregation) {Joiner};
			\node[artifact, fill=white, left=8mm of split.west] (prog) {};
			\node[below = 0.5mm of prog.south, anchor=north,align=left] (prog-label) {Program,\\Property};	
			\node[artifact, fill both halfs={colorLower}{colorUpper},right=\hdiffWider of aggregation.east](res) {};
			\node[below = 0.5mm of res.south, anchor=north] (res-label) {Result};			
			\node[above right= 0.75cm and -0.1cm of ra1, anchor=north] (orchestration) {\textbf{Ranged Program Analysis with Work Stealing}};

			\path (prog) edge[flow]  (split);			
			\path (split) edge[flow,bend left=10,] node[verdict, fill=colorUpper] {Range\\[-1mm] [$\inducedTau, \noUpperBound$]}   (ra1.west);
		
			\path (split) edge[flow,bend right=10,]node[verdict, fill=colorLower] {Range\\[-1mm][$\noLowerBound, \inducedTau$] } (ra4.west);		
			\path ($(ra1.south east)$) edge[flow,bend right=15,pos=0.4] node[artifact,fill upper half=colorUpper,label={[xshift=-1.1cm, yshift=-1.0cm,align=center]Partial\\[-1mm]Result}] {} ($(aggregation.north west)+ (0,-0.2)$);
			\path (ra1) edge[flow]node[verdict, fill=colorLower] {Range\\[-1mm][$\noLowerBound, \inducedTau$] } (ra1new);		
			\path (ra1new) edge[flow,pos=0.2, out=270, in=00] node[artifact,fill lower half=colorLower, xshift=-0.2cm, yshift=-0.09cm,label={[xshift=-1.1cm, yshift=-1.0cm,align=center]Partial\\[-1mm]Result}] {} ($(aggregation.north east)+ (0,-0.2)$);
			\path (aggregation) edge[flow, bend right=0]  (res);
			\path ($(aggregation.south west)$) edge[flow, dotted, draw=gray]  node[rotate=38,yshift=2.5mm]{\textcolor{gray}{terminate}} (ra4);
			\begin{pgfonlayer}{bg}
				
				\draw[black, draw, fill=white] ($(split.north west) + (-.3, 3.1)$) rectangle ($(aggregation.south east) + (3.6,-2.5)$) ;
				\path ($(prog.east) + (-.2,+.3) $) edge[flow,bend left=25,]  ($(ra1.west) + (0,+.35) $);
				\path ($(prog.east) + (0,-.3) $) edge[flow,bend right=25, ] ($(ra4.west) + (0,-.35) $);
			\end{pgfonlayer}

	\end{tikzpicture}}
	
	\caption{Ranged analysis with work stealing, where \textsf{Ranged Analysis 1} completes the verification of both ranges
		\label{fig:workstealingRA1Only}}	
\end{figure}
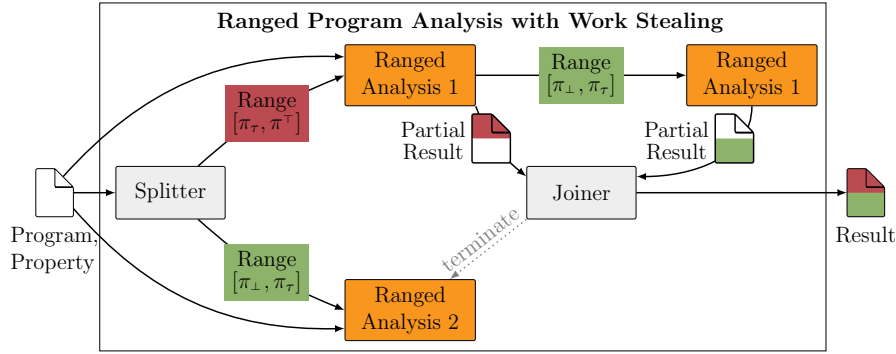
\begin{figure}[t]
	\centering
	\newcommand{\hdiff}{12mm}
	\newcommand{\hdiffWide}{25mm}
	\newcommand{\hdiffWider}{40mm}
	\newcommand{\vdiff}{15mm}
	\scalebox{0.7}{
		\begin{tikzpicture}
			\pgfdeclarelayer{bg}
			\pgfsetlayers{bg,main}
			\node[ concept_small] (split) {Splitter};
			\node[right= \hdiffWide of split.east, draw=none, minimum width=2cm] (ra2){};
			\node[ ra_small, above=\vdiff of ra2] (ra1) {Ranged Analysis 1};
			\node[ ra_small, right=\hdiffWider of ra1] (ra1new) {Ranged Analysis 1};
			\node[ ra_small, below=\vdiff of ra2] (ra4) {Ranged Analysis 2};
			\node[ concept_small, right = \hdiff of ra2.east] (aggregation) {Joiner};
			\node[artifact, fill=white, left=8mm of split.west] (prog) {};
			\node[below = 0.5mm of prog.south, anchor=north,align=left] (prog-label) {Program,\\Property};	
			\node[artifact, fill both halfs={colorLower}{colorUpper},right=\hdiffWider of aggregation.east](res) {};
			\node[below = 0.5mm of res.south, anchor=north] (res-label) {Result};			
			\node[above right= 0.75cm and -0.1cm of ra1, anchor=north] (orchestration) {\textbf{Ranged Program Analysis with Work Stealing}};	
			
			\path (prog) edge[flow]  (split);			
			\path (split) edge[flow,bend left=10,] node[verdict, fill=colorUpper] {Range\\[-1mm] [$\inducedTau, \noUpperBound$]}   (ra1.west);
			\path (split) edge[flow,bend right=10,]node[verdict, fill=colorLower] {Range\\[-1mm][$\noLowerBound, \inducedTau$] } (ra4.west);		
			\path ($(ra1.south east)$) edge[flow,bend right=15,pos=0.4] node[artifact,fill upper half=colorUpper,label={[xshift=-1.1cm, yshift=-1.0cm,align=center]Partial\\[-1mm]Result}] {} ($(aggregation.north west)+ (0,-0.2)$);
			\path (ra1) edge[flow]node[verdict, fill=colorLower] {Range\\[-1mm][$\noLowerBound, \inducedTau$] } (ra1new);		
			\path ($(ra4.north east)$) edge[flow,bend left=15,pos=0.4] node[artifact,fill lower half=colorLower,label={[xshift=-1.1cm, yshift=-1.0cm,align=center]Partial\\[-1mm]Result}] {} ($(aggregation.south west)+ (0, 0.2)$);
			\path (aggregation) edge[flow, bend right=0]  (res);
			\path ($(aggregation.north east)$) edge[flow, dotted, draw=gray]  node[rotate=38,yshift=-2.0mm]{\textcolor{gray}{terminate}} (ra1new);
			\begin{pgfonlayer}{bg}
				
				\draw[black, draw, fill=white] ($(split.north west) + (-.3, 3.1)$) rectangle ($(aggregation.south east) + (3.6,-2.5)$) ;\path ($(prog.east) + (-.2,+.3) $) edge[flow,bend left=25,]  ($(ra1.west) + (0,+.35) $);
				\path ($(prog.east) + (0,-.3) $) edge[flow,bend right=25, ] ($(ra4.west) + (0,-.35) $);
			\end{pgfonlayer}

	\end{tikzpicture}}
	
	\caption{Ranged analysis with work stealing, where both \textsf{Ranged Analysis 1} and \textsf{Ranged Analysis 2} complete the verification of a range \label{fig:workstealingBoth}}	
\end{figure}
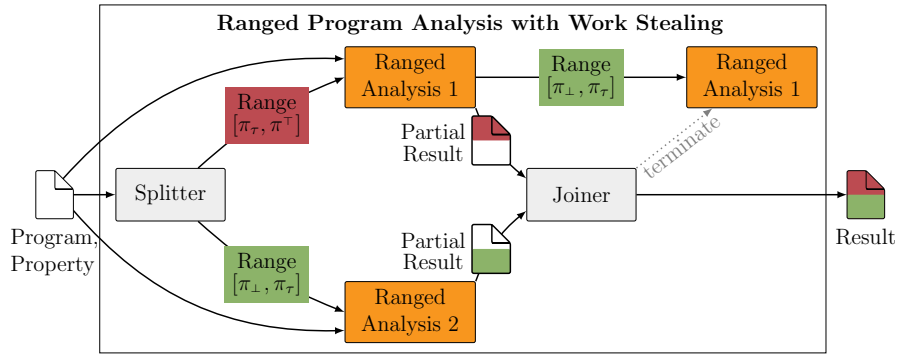

As the assignment of program analyses and ranges is fixed for all tasks, we propose using \emph{work stealing} to avoid the situation that the ranged program analysis fails at solving a task, although at least one of the used ranged analyses can solve the task.
The idea of work stealing is simple but efficient:
The ranged analysis that first finishes the analysis of its assigned range is restarted on the other range unless it already found a violation and, thus, already determined the analysis result.
Thereby, the other range is analyzed by both ranged analyses in parallel, until one computes a result for the range.
We depict work stealing in case the analysis of the range $[\inducedTau, \noUpperBound]$ is completed first in \cref{fig:workstealingRA1Only} and  \cref{fig:workstealingBoth}.
In both scenarios, \textsf{Ranged Analysis 1} is started for a second time and analyzes the range $[\noLowerBound, \inducedTau]$.
In \cref{fig:workstealingRA1Only}, we depict the situation that \textsf{Ranged Analysis 1} also finishes for 
the range $[\noLowerBound, \inducedTau]$, i.e. it successfully analyzes both ranges, and \textsf{Ranged Analysis 2} is terminated.
\Cref{fig:workstealingBoth} depicts the opposite situation, where \textsf{Ranged Analysis 2} finishes the analysis of the range $[\noLowerBound, \inducedTau]$ before \textsf{Ranged Analysis 1} and \textsf{Ranged Analysis 1} is terminated.
Note that work stealing offers a simple but efficient way to reduce the risk of the ranged program analysis failing, especially in cases where one ranged analysis used can solve the task.
Conceptually, work stealing also allows for a precision reuse~\cite{DBLP:conf/sigsoft/BeyerLNSW13}, as the analysis that is restarted on the second range can reuse the precision from its first iteration, e.g., the set of discovered predicates.
As a trade-off, work stealing increases resource consumption, as there may run three analyses in total (but at most two in parallel). 


\section{Implementation}\label{sec:implementation}
To show the advantages of ranged program analysis and to evaluate the two novel techniques of witness joining and work stealing,
we implemented ranged program analysis and the two novel extensions within \coveriteam~\cite{CoVeriTeam}, a tool for specifying compositions of verification tools.

\inlineheadingbf{Ranged Analyses within \cpachecker}
At first, we implement the composition of range reduction from \cref{sec:range-reduction} with an arbitrary configurable program analysis in the tool \cpachecker~\cite{CPAchecker}, making use of the existing composite pattern.
    More precisely, we build a novel range reduction analysis within \cpachecker, which combines the transfer relations of lower bound CPA and upper bound CPA as formalized in \cref{sec:range-reduction}.
    Hence, it reuses elements from the value analysis, especially from the transfer relation, which are already implemented within \cpachecker.
Our implementation is open-source and published in the \cpachecker repository.
To ease reproducibility and practical applicability, we included the implementation also in our artifact archived at \zenodo~\cite{artifact}.

\inlineheadingbf{Ranged Analyses not using \cpachecker}

For the use of non-CPA-based analyses, we refer to~\cite{RangedAnalysisSEFM}, where a ranged program analysis based on instrumentation is presented.
It allows for using any off-the-shelf analysis.
Instead of composing the analysis with a range reduction, a so-called range program is generated.
For this, reduction~\cite{DBLP:conf/icse/BeyerJLW18} and program instrumentation are used.

\inlineheadingbf{Ranged Program Analysis}
The composition of ranged program analyses from \cref{sec:combined-range-analysis} is implemented within the tool \coveriteam~\cite{CoVeriTeam}.
The tool allows building parallel and sequential compositions making use of existing program analyses, especially those from \cpachecker.
To be able to use work stealing, i.e., to keep track of which program range is already successfully analyzed, we build a novel \emph{ranged program analysis component} within \coveriteam.
Its task is to (1) orchestrate the composition of ranged analyses, especially for work stealing, and (2) the aggregation of the partial results.
The algorithm follows the description in \cref{fig:overview-ranged-analysis} and \cref{sec:work-stealing} and receives as inputs the two ranged analyses \textsf{Ranged Analysis 1} and \textsf{Ranged Analysis 2}, and a splitter.
It first uses the splitter to generate the ranges.
If the splitter fails, e.g., \LBThree\xspace cannot compute a test case, when the program does not contain a loop, we execute  \textsf{Ranged Analysis 1} on the interval $[\noLowerBound, \noUpperBound]$.
We extended \coveriteam by adding ranged analyses as a new type of verifier.
Thus, it executes the two ranged analyses in parallel using the execution mechanism of \coveriteam.
The algorithm collects the verdicts and witnesses computed by the ranged analyses and employs, if configured, work stealing.
Moreover, the witnesses are joined.
Therefore, the partial verification witnesses generated by \textsf{Ranged Analysis~1} and \textsf{Ranged Analysis 2} are collected first.
For the witnesses join, we implemented \cref{alg:joinWit} as a standalone component within \cpachecker, which is executed by \coveriteam.
Our implementation is open-source and publicly available in the repository of \coveriteam as well as in the artifact~\cite{artifact}.

\inlineheadingbf{Generating Ranges}
We implemented the splitters \LBThree that is described in \cref{sec:gen-ranges} as standalone components within \cpachecker.
It generates test cases in the standardized XML-based \testcomp test-case format\footnote{\url{https://gitlab.com/sosy-lab/test-comp/test-format/blob/testcomp23/doc/Format.md}}.
Therefore, it builds the program execution tree demand-driven, meaning that only the path containing three loop unrollings is generated.
Using the path formula, a representation of the path using predicates, an \tool{SMT}-solver is called to find satisfiable assignments for the program input variables.
These assignments are then used within the test case that is returned.


\section{Evaluation}
The evaluation of ranged symbolic execution conducted by Siddiqui and Khurshid~\cite{Sym-Range} focuses on the performance increases gained for symbolic execution.
	We showed in~\cite{RangedAnalysisFASE} that ranged program analysis can increase the overall performance for finding property violations of symbolic execution by \num{7}\%.
	In addition, we compared the performance of the four different splitters introduced in \cref{sec:gen-ranges}, whereas \LBThree leads to the best overall performance.
	Moreover, the real overall execution time that is needed to solve large or complex tasks can be reduced by using ranged program analysis by up to 20\%.
	When using ranged program analysis with BMC, it is observed that tasks that could not be solved using a single BMC instance can be solved when using BMC within ranged program analysis.
	We also evaluated one combination or ranged program analysis that uses two different analysis techniques, namely symbolic execution and predicate abstraction.
	We observed that the performance increases for symbolic execution, the one with the lower performance.

In the following, we are interested in a more systematic analysis of the effect of using ranged program analysis.
The overall goal of the evaluation is to investigate, to which extent using ranged program analysis can increase the performance (i.e. effectiveness and efficiency) of analyses like predicate abstraction or symbolic execution.
Hence, we compare ranged program analysis with the basic analysis employed.
First, we want to investigate whether using two instances of the same analysis within ranged program analysis pays off.
Thereafter, we study the effect of combining two conceptually different analyses and also analyze work stealing.
Lastly, we study the question of whether our ranged program analysis is still capable of generating valid correctness witnesses by using the novel technique for witness joining.
To this end, we studied the following research questions:
\begin{description}
	\item[RQ1] Can ranged program analysis increase the efficiency and effectiveness of CPA-based analyses?
	\item[RQ2] Can a combination of different analyses within a ranged program analysis also increase the efficiency and effectiveness of the analyses?
	\item[RQ3] Does work stealing pay off?
	\item[RQ4] Does ranged program analysis using witness joining generate valid verification witnesses?
\end{description}

\subsection{Employed CPA-based Analysis.}
For the evaluation, we used combinations of four existing CPA-based program analyses, called \emph{basic analyses}, within the ranged analysis, briefly introduced next.

\inlineheadingbf{Value Analysis}
Value Analysis~\cite{ValueAnalysis} (sometimes called constant propagation or explicit state model checking) tries to compute the concrete values for variables occurring in the program.
For this, the exact value of each variable from a set (called precision) is tracked along each program path.
Information for locations reachable via multiple program paths is not merged.
The employed configuration does not track all variable values but uses counterexample-guided abstraction refinement (CEGAR)~\cite{ClarkeEtAl:2000} to iteratively refine the precision until the program can either be proven safe or a violation is detected. 

\inlineheadingbf{Symbolic Execution}
Symbolic execution~\cite{King76}  uses symbolic inputs instead of concrete values for analyzing program paths.
A state is a pair of a symbolic store, describing variable values by formulae using the symbolic inputs, and a path condition, tracking the executability of the explored path.
Each operation updates the symbolic store.
At a branching point, the current path condition is extended using the branching condition evaluated with the symbolic values.
The exploration of a path is stopped when the path condition becomes unsatisfiable or a program end is reached.

\inlineheadingbf{Predicate Abstraction}
The predicate abstraction used is the standard one in \cpachecker, configured to compute an abstraction of the program using predicates with adjustable block encoding~\cite{ABE}, abstracting only at loop heads.
The predicates used for the analysis are determined by CEGAR~\cite{ClarkeEtAl:2000}, lazy refinement~\cite{LazyAbstraction}, and inter\-polation~\cite{AbstractionsFromProofs}.

\inlineheadingbf{Bounded Model Checking}
The configuration of bounded model checking~(BMC) employed is iterative, meaning that the $k$-th iteration inspects the behavior of the CFA by unrolling each loop to a bound of $k$ iterations.
If no property violation can be found, $k$ is increased.
For inspecting the behavior of the program, BMC encodes the unrolled CFA and the property as a boolean formula using the unified SMT-based \mbox{approach} for software verification~\cite{AlgorithmComparison-JAR}. 
Hereafter, the satisfiability of the formula that encodes the program up to the bound $k$ is checked to detect property violations.
In case that $k$ is large enough that the full program is encoded, and no property violation is detected, the program is safe.

\smallskip
For the evaluation, we build different configurations of ranged program analysis using the four basic analyses.
We use the abbreviations \Value for value analysis, \SymbExec for symbolic execution, \Pred for predicate abstraction and \BMC for BMC.
A ranged program analysis that uses value analysis for the range $[\noLowerBound,\pi_\tau]$ and predicate abstraction for $[\pi_\tau,\noUpperBound]$ for some computed test input $\tau$ is denoted by \RAValuePred.
We indicate the use of work stealing by using the abbreviation \WorkStealing for the configuration.
For our evaluation, we will only make use of the splitter \LBThree, as the experimental evaluation in~\cite{RangedAnalysisFASE} has shown that it leads to the best performance for ranged program analysis.

\subsection{Evaluation Setup}
All experiments were run on machines with an Intel Xeon E3-1230 v5 @ 3.40 GHz (8~cores), \num{33}\,GB of memory, and Ubuntu~22.04~LTS with Linux kernel~5.15.0.
To increase the reproducibility of our experiments, we used \benchexec~\cite{Benchmarking-STTT} for the execution.\\
We conducted the experiments on a subset of the \svbenchmarks used in the \svcomp for the property reachability and all experiments were conducted once.
The subset contains in total \num{7244}~C-tasks from all sub-categories of the \svcomp category \texttt{ReachSafety} and \texttt{ConcurrencySafety} dealing with the reachability of error labels~\cite{svbenchmarks23}, whereof \num{4242}~tasks fulfill the property and the other \num{3002} contain a property violation.
In a so-called \emph{verification run} a tool is given one the \num{7244}~C-tasks, containing a program and a specification, and is asked to either compute a proof (in case the program fulfills the specification) or to raise an alarm (if it violates the specification). 
We limit the available resources to a total of \num{15}\,min CPU~time on \num{4}~CPU~cores and \num{15}\,GB of memory, yielding a setup that is comparable to the one used in \svcomp.
To ensure a comparison on equal ground, we use the same resource limitations for each tool.
This means that the basic analyses have the same resources available as the ranged program analyses.
As we employ two ranged analyses within a ranged program analysis, the available resources are split between the two ranged analyses.\footnote{
	We evaluated configurations using three ranged analyses in parallel in~\cite{RangedAnalysisFASE}.}

All data collected is available in our supplementary artifact~\cite{artifact}.

%
%
%
%
%
%
%
%
%
%
\subsection{RQ 1: Ranged Program Analysis Using Two Identical Analyses}\label{sec:RQ1}
\inlineheadingbf{Evaluation Plan}
To analyze the performance of the four basic analyses within ranged program analysis, we compare the effectiveness (number of tasks solved) and efficiency (time taken to solve a task) for the ranged program analysis against the basic analysis running standalone.
For efficiency, we focus on the (real) time taken overall to solve the task (called \textit{wall~time}), as we want to take the advantages of the parallelization employed within the ranged program analysis into account.
Note that we limit the resources for the overall configuration, hence the available resources are shared between both ranged analyses running in parallel within the ranged program analysis.
To achieve a fair comparison, we also executed the basic analyses in \coveriteam,
where we build a simple configuration that directly calls  \cpachecker running the basic analyses.

\begin{table}[t]
	\caption{
		Number of correct and incorrect verdicts reported by the four basic analyses and the combination of ranged program analysis.		
			The column \textit{par.only} (parallel only) contains the number of tasks that are solved by the parallel combinations, either ranged program analysis (\textsc{Ra}) or ranged program analysis using work stealing (\textsc{Ws}) but not by the basic analysis.
			If a parallel combination uses two different basic analyses, the tasks reported are not solved by any of the two basic analyses.
	}
	\centering
	\setlength{\tabcolsep}{4pt}	
	\begin{tabular}{l S[table-format=4.0]S[table-format=4.0] S[table-format=4.0]  S[table-format=3.0] @{\hspace{1.5em}} S[table-format=1.0] S[table-format=2.0] }
		\toprule
		                      & \multicolumn{4}{c}{\textbf{correct} } & \multicolumn{2}{c}{\textbf{incorrect} }                                                                                                                                                                                         \\
		                      & {\textbf{overall}}                    & {\textbf{proof}}                        & {\textbf{alarm}}          & \textbf{par.only}                & {\textbf{proof}}           & {\textbf{alarm}}           \\
		\midrule
		\textbf{\SymbExec}    & \correctOverallSE                     & \correctProofSE                         & \correctAlarmSE           & {-}                            & \incorrectProofSE          & \incorrectAlarmSE          \\
		\textbf{\Value}       & \correctOverallValue                  & \correctProofValue                      & \correctAlarmValue        & {-}                            & \incorrectProofValue       & \incorrectAlarmValue       \\
		\textbf{\Pred}        & \correctOverallPred                   & \correctProofPred                       & \correctAlarmPred         & {-}                            & \incorrectProofPred        & \incorrectAlarmPred        \\
		\textbf{\BMC}         & \correctOverallBMC                    & \correctProofBMC                        & \correctAlarmBMC          & {-}                            & \incorrectProofBMC         & \incorrectAlarmBMC         \\
		\midrule
		\textbf{\RASE}        & \correctOverallRASE                   & \correctProofRASE                       & \correctAlarmRASE         & \additionalVsFirstRASE       & \incorrectProofRASE        & \incorrectAlarmRASE        \\
		\textbf{\RAValue}     & \correctOverallRAValue                & \correctProofRAValue                    & \correctAlarmRAValue      & \additionalVsFirstRAValue    & \incorrectProofRAValue     & \incorrectAlarmRAValue     \\
		\textbf{\RAPred}      & \correctOverallRAPred                 & \correctProofRAPred                     & \correctAlarmRAPred       & \additionalVsFirstRAPred     & \incorrectProofRAPred      & \incorrectAlarmRAPred      \\
		\textbf{\RABMC}       & \correctOverallRABMC                  & \correctProofRABMC                      & \correctAlarmRABMC        & \additionalVsFirstRABMC      & \incorrectProofRABMC       & \incorrectAlarmRABMC       \\
		\midrule
		\textbf{\RASEPred}    & \correctOverallRASEPred               & \correctProofRASEPred                   & \correctAlarmRASEPred     & \additionalVsBothRASEPred    & \incorrectProofRASEPred    & \incorrectAlarmRASEPred    \\
		\textbf{\RAValuePred} & \correctOverallRAValuePred            & \correctProofRAValuePred                & \correctAlarmRAValuePred  & \additionalVsBothRAValuePred & \incorrectProofRAValuePred & \incorrectAlarmRAValuePred \\
		\textbf{\RABMCPred}   & \correctOverallRABMCPred              & \correctProofRABMCPred                  & \correctAlarmRABMCPred    & \additionalVsBothRABMCPred   & \incorrectProofRABMCPred   & \incorrectAlarmRABMCPred   \\
		\midrule
		\textbf{\WSSEPred}    & \correctOverallWSSEPred               & \correctProofWSSEPred                   & \correctAlarmWSSEPred     & \additionalVsBothWSSEPred    & \incorrectProofWSSEPred    & \incorrectAlarmWSSEPred    \\
		\textbf{\WSValuePred} & \correctOverallWSValuePred            & \correctProofWSValuePred                & \correctAlarmWSValuePred  & \additionalVsBothWSValuePred & \incorrectProofWSValuePred & \incorrectAlarmWSValuePred \\
		\textbf{\WSBMCPred}   & \correctOverallWSBMCPred              & \correctProofWSBMCPred                  & \correctAlarmWSBMCPred    & \additionalVsBothWSBMCPred   & \incorrectProofWSBMCPred   & \incorrectAlarmWSBMCPred   \\
		\bottomrule
	\end{tabular}
	\label{tab:RQ1}
	\vspace{-5mm}
\end{table}

\inlineheadingbf{Effectiveness}
We report on the obtained results for RQ1 in the first two blocks of \cref{tab:RQ1}.
	Each row contains the number of overall correctly solved tasks, the number of correctly computed proofs and alarms, and the number of tasks solved only by the parallel combinations but not by the basic analyses used in the configurations. 
Moreover, \cref{tab:RQ1} contains the number of incorrectly computed proofs and alarms.
When analyzing the effectiveness of ranged program analysis for symbolic execution, we observe that \RASE correctly solves in total \num{\correctOverallRASE} tasks and \SymbExec in total \num{\correctOverallSE} tasks, meaning that \RASE can solve in total \num{\moreSolvedVsFirstRASE} tasks more than the basic symbolic execution, an increase of \num{\moreSolvedPercentageVsFirstRASE}\%.
Moreover, there are \num{\additionalVsFirstRASE} tasks that can be solved only using ranged program analysis but not \SymbExec.
In all \num{\additionalVsFirstRASE} cases, \RASE computed a valid alarm.
The number of reported proofs is nearly constant, as \SymbExec and \RASE both have to check all paths in the program leading to a property violation for infeasibility.

For value analysis, we see that \Value solves in total \num{\correctOverallValue} tasks and \RAValue \num{\correctOverallRAValue} tasks, meaning that \RAValue solves in total \num{\moreSolvedVsFirstRAValue} tasks more than the basic analysis, an increase of \num{\moreSolvedPercentageVsFirstRAValue}\%. 
Again, we observe that \RAValue can solve \num{\additionalVsFirstRAValue} tasks that are not solved by value analysis, nearly exclusively containing correctly raised alarms, as the value analysis is also a path-based analysis that does not apply abstraction mechanisms.

\newcommand{\scalingFactor}{0.45}
\newcommand{\scalingFactorVenn}{0.45}
\begin{figure}[t]
	\begin{minipage}{\linewidth}
		\captionsetup[subfigure]{justification=centering}
		\hfil
		\subfloat[\SymbExec\label{fig:scatterSymbExecRQ1}]{\scalebox{\scalingFactor}{
				\includegraphics[trim=0.5cm 0.5cm 0.4cm 1cm]{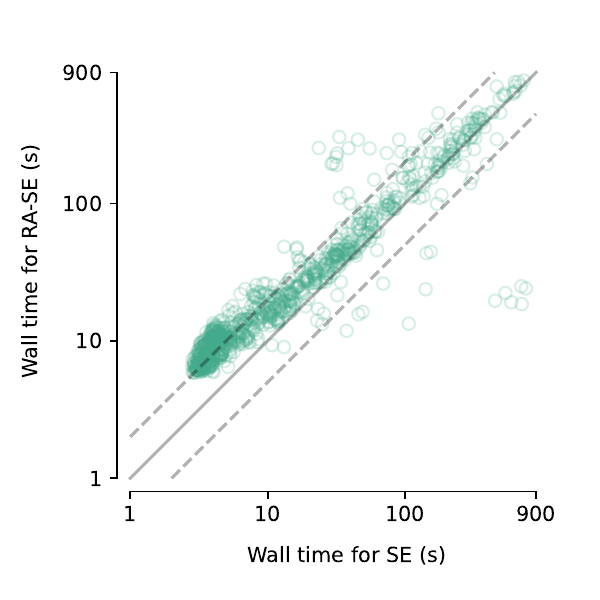}}
		}\hfil
		\subfloat[\Value\label{fig:scatterValueRQ1}]{\scalebox{\scalingFactor}{
				\includegraphics[trim=0.5cm 0.5cm 0.4cm 1cm]{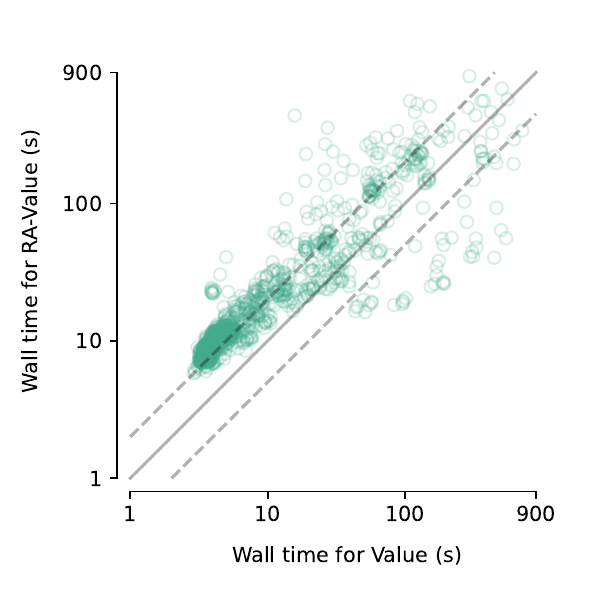}}
		}
		\hfil
		\subfloat[\Pred\label{fig:scatterPredRQ1}]{\scalebox{\scalingFactor}{
				\includegraphics[trim=0.5cm 0.5cm 0.4cm 1cm]{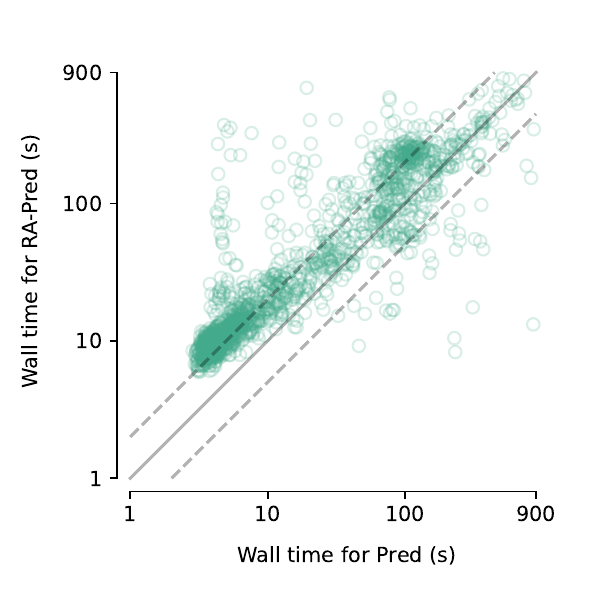}}
		}\hfil
		\subfloat[\BMC\label{fig:scatterBMCRQ1}]{\scalebox{\scalingFactor}{
				\includegraphics[trim=0.5cm 0.5cm 0.4cm 1cm]{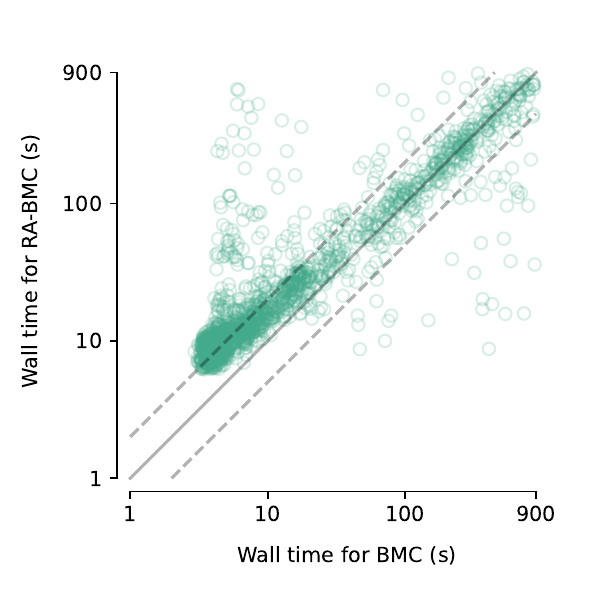}}
		}
		\vspace*{-3mm}
		\caption{Scatter plots comparing the wall time of the basic analyses and ranged program analysis using two instances of the same analysis. \label{fig:scatter-rq1}}
	\end{minipage}
	\vspace*{-5mm}
\end{figure}
	Having a look at predicate abstraction and BMC, we again notice that there are tasks that are solved by the ranged program analyses only and not by the basic analyses, 
	namely \num{\additionalVsFirstRAPred} for \RAPred and \num{\additionalVsFirstRABMC} for \RABMC.
	In contrast to value analysis and symbolic execution, the combination using either two instances of BMC or predicate abstraction within the ranged program analysis does not increase but slightly decreases the overall number of correctly solved tasks:
	\RAPred can solve overall \num{\correctOverallRAPred} tasks and the basic analysis \Pred \num{\correctOverallPred} tasks, a decrease of \num{\moreSolvedVsFirstRAValue}.
Similarly, \RABMC solves \num{\correctOverallRABMC} tasks and \BMC \num{\correctOverallBMC} tasks, a decrease of \num{\moreSolvedVsFirstRAValue}.
This decrease is most likely caused by the fact that both analyses consider more than one path of the program, either through abstraction or full program encoding.

Having a look at the number of incorrectly computed results, we observe that \RASE, \RAValue, and \RAPred raise more incorrect alarms compared to the basic analyses.
To validate that the additional incorrect alarms are not caused by an error in the implementation of the ranged program analysis, we analyzed a set of randomly selected tasks and restricted the basic analyses to explore only the path encoded in the reported violation.
For all tasks, the basic analyses now also raised the alarms instead of running into a timeout.

\inlineheadingbf{Efficiency}
\begin{figure}[h]
	\begin{minipage}{\linewidth}
		\captionsetup[subfigure]{justification=centering}\hfil
		\subfloat[\SymbExec\label{fig:app:increaseSeRQ1}]{\scalebox{0.5}{
				\includegraphics[trim=0.5cm 0.2cm 0.4cm 0.2cm]{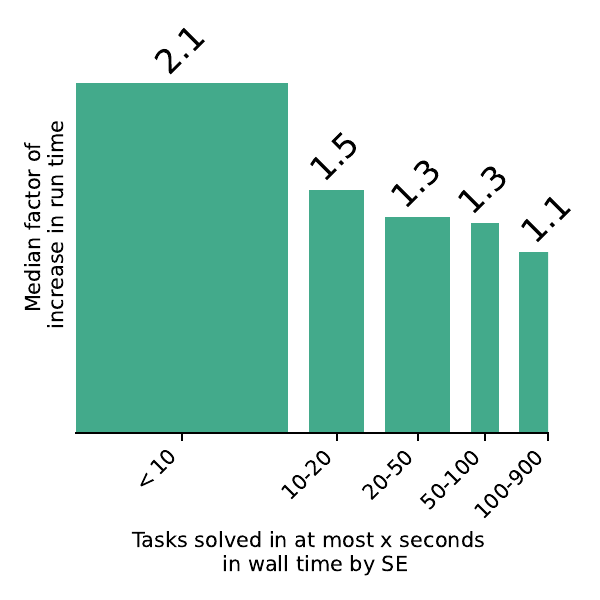}}
		} \hfil
		\subfloat[\Value\label{fig:app:increaseValueRQ1}]{\scalebox{0.5}{
				\includegraphics[trim=0.5cm 0.2cm 0.4cm 0.2cm]{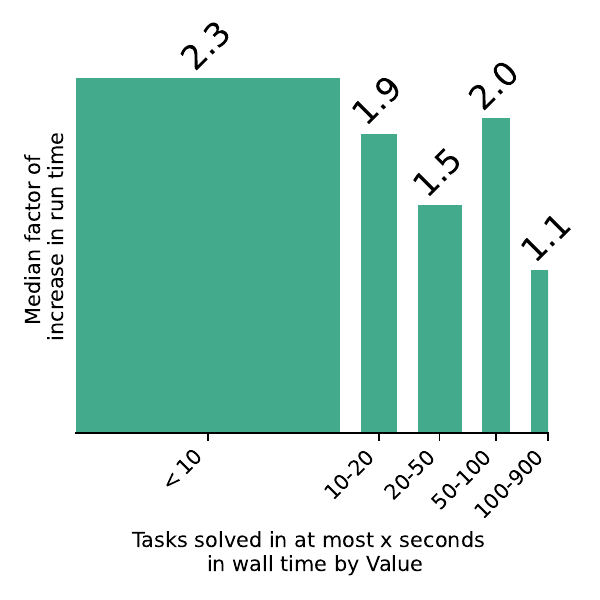}}
		}

		\hfil
		\subfloat[\Pred\label{fig:app:increasePredRQ1}]{\scalebox{0.5}{
				\includegraphics[trim=0.5cm 0.2cm 0.4cm 0.2cm]{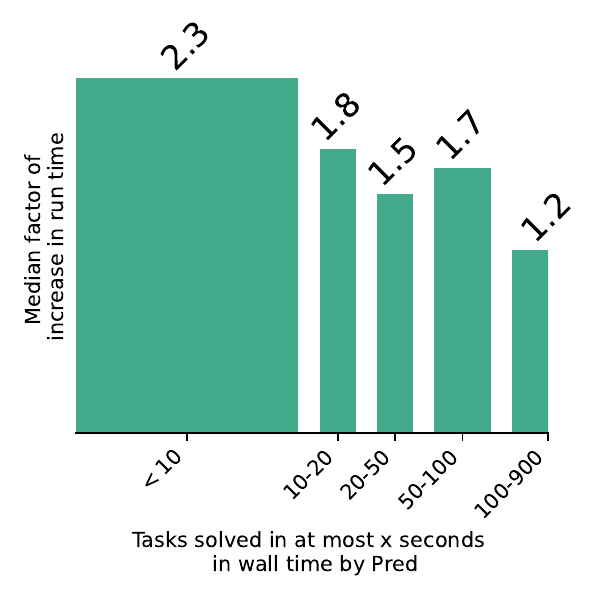}}
		}\hfil
		\subfloat[\BMC\label{fig:app:increaseBMCRQ1}]{\scalebox{0.5}{
				\includegraphics[trim=0.5cm 0.2cm 0.4cm 0.2cm]{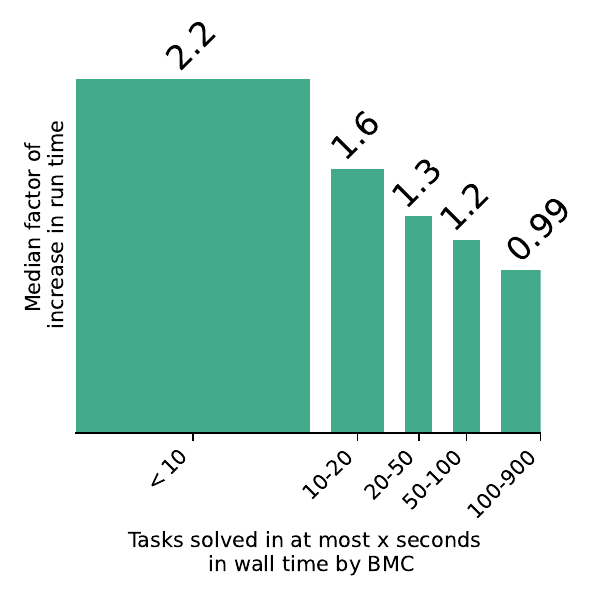}}
		}
		\hfill
		\caption{Median wall~time increase of the default analyses compared to the ranged program analyses. The width of the bars corresponds to the number of tasks solved by the default analysis in that time interval. \label{fig:app:increaseRQ1}}
	\end{minipage}
\end{figure}
For analyzing the efficiency of the ranged program analysis, we present in \cref{fig:scatter-rq1} scatter plots comparing the wall~time taken by the tools in a log-scale, in case both tools can solve the task and a range is generated.
On the x-axis, we depict the time taken by the basic analyses, on the y-axis the wall~time of the ranged program analyses.
A point $(x,y)$ in the scatter plot means that the basic analysis solved the tasks in $x$ seconds wall~time and the ranged program analysis in $y$ seconds.
For points below the diagonal, the ranged program analysis can solve the tasks faster than the basic analysis, otherwise, it is slower.
The dashed lines indicate, where tools are twice as fast / twice as slow as the other.
At first glance, we realize that the ranged program analyses have a somewhat similar runtime to the basic analyses.

For a more detailed comparison of the wall~time taken, we use \cref{fig:app:increaseRQ1}.
We depict the time taken by basic analyses on the x-axis, grouped in intervals.
The height of each bar represents the median value of the time increase for ranged program analysis, where the width of each bar corresponds to the number of tasks that are solved by the basic analyses in the interval.
For all four basic analyses, we make two observations:
Firstly, the ranged program analysis takes approximately twice as long to compute a solution for tasks that can be solved fast (in less than ten seconds) by the basic analysis.
Secondly, we observe that for complex tasks the ranged program analysis becomes faster w.r.t. the basic analysis, although there are some outliers, such that we notice a comparable wall~time of basic analysis and ranged program analysis for the complex tasks taking more than \num{100}s to solve.

\begin{wrapfigure}{r}{0.45\textwidth}
		\subfloat[CPU~time\label{fig:cpu-lbthree}]{\scalebox{0.5}{
				\includegraphics[trim=0.5cm 0.6cm 0.4cm 2.2cm]{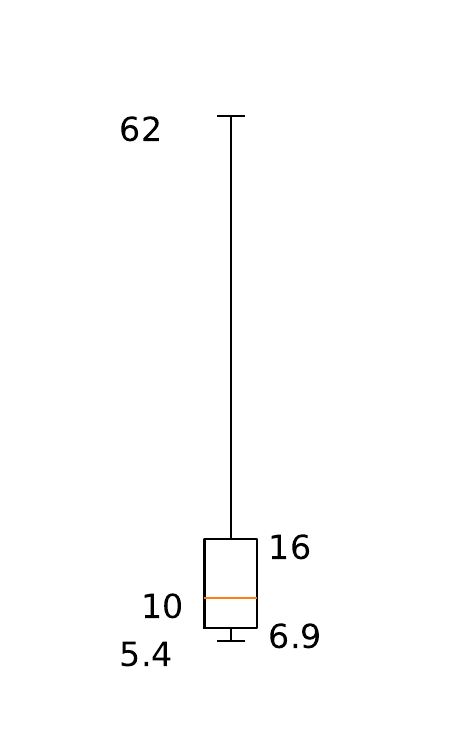}}
		}		
		\subfloat[Wall~time \label{fig:wall-lbthree}]{\scalebox{0.5}{
				\includegraphics[trim=0.5cm 0.6cm 0.4cm 3.3cm]{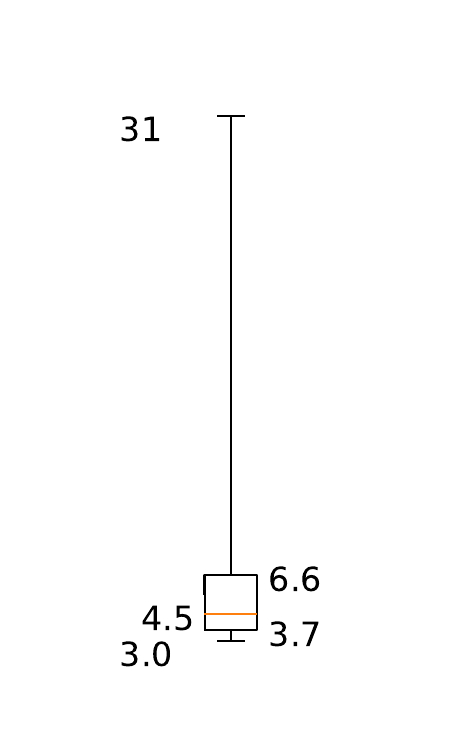}}
		}
		\caption{Boxplot for CPU~time and wall~time to compute a test case using \LBThree. \label{fig:time-lbthree}}
	\end{wrapfigure}
To analyze why the ranged program analyses need more overall time to compute a solution compared to the basic analyses, especially for tasks that can be solved fast, we have a more detailed look at the individual steps:
We analyze the time needed to compute the ranges.
Therefore, we measured the time taken by \LBThree to compute a test case separately for all tasks and depicted the results as boxplots in \cref{fig:time-lbthree}.
Therein, we depict the CPU~time taken by \LBThree in \cref{fig:cpu-lbthree} and the wall~time in \cref{fig:wall-lbthree}.
In the median, \LBThree needs \num{10} seconds CPU~time and overall \num{4.5} seconds to compute the bounds, visualized using the orange lines.
The boxes contain the median values for \num{25}\% resp. \num{75}\% of the tasks.
Hence, \num{75}\% of all bounds are computed within \num{16} seconds CPU and \num{6.6} seconds wall time.
The whisker contains \num{99}\% of all tasks, hence there are only very few cases where \LBThree needs more than \num{62} resp. \num{31} seconds.

As computing the test cases takes in the median \num{10} seconds CPU~time and \num{4.5} seconds wall time, this additional computational effort is a main factor for the time increase.
As the time taken to compute a range is approximately the same for all tasks, it influences the overall wall time for complex tasks only to a small extent.
For the more complex tasks, the advantages of ranged program analysis, namely analyzing different parts of the program in parallel, become visible.
Unfortunately, the additional overhead caused by using ranged program analysis exceeds the advantages of parallelization, as the ranged program analysis is nearly as fast as the basic analysis for complex tasks, but not faster.

\inlineheadingbf{Summary}
Using the same analysis within ranged program analysis increases the effectiveness of the path-based analyses symbolic execution and value analysis.
Each instance of the ranged program analyses can solve tasks that are not solved by the basic analyses used.
The efficiency is comparable for larger and more complex tasks, and computing the ranges yields a constant overhead.
%
%
%
%
%
%
%
%
%
%
\subsection{RQ 2: Using Different Analyses Within Ranged Program Analysis}\label{sec:RQ2}
\inlineheadingbf{Evaluation Plan}
To investigate whether other analysis combinations benefit from a composition of ranged analyses, we evaluated three different compositions:
We combine predicate abstraction with symbolic execution, value analysis, and BMC.
Thereby, we get the three configurations \RASEPred,  \RAValuePred, and \RABMCPred for ranged program analysis.
	In the three configurations, we combine predicate abstraction, a technique that can effectively compute proofs with techniques that are good at detecting property violations. 
We compare the ranged program analyses to the two basic analyses used and are again interested in effectiveness and efficiency.

\inlineheadingbf{Effectiveness}
To evaluate the effectiveness of the combinations using two different analyses within ranged program analysis,
we have to compare the first and the third block of \Cref{tab:RQ1}.
The column \textit{par.only} now contains the tasks that neither of the two basic analyses can solve but the ranged program analysis, whereas the other columns are as explained in RQ~1.

	For \RASEPred, the ranged program analysis can solve in total \num{\correctOverallRASEPred} tasks, that are \num{\moreSolvedVsFirstRASEPred} tasks more than symbolic execution can solve standalone (\num{\correctOverallSE}).
	Unfortunately, \RASEPred does not solve more tasks than the predicate abstraction can solve standalone (\num{\correctOverallPred}).
For most of these \num{\moreSolvedVsOtherRASEPred} tasks, the predicate abstraction successfully verifies the assigned range, but the symbolic execution fails to solve its part (range) of the task.
We can also observe that the ranged program analysis can solve \num{\additionalVsBothRASEPred} tasks that neither the symbolic execution nor the predicate abstraction can solve.

\begin{figure}[t]
	\begin{minipage}{\linewidth}
		\captionsetup[subfigure]{justification=centering}
		\hfil
		\subfloat[\SymbExec and \RASEPred\label{fig:scatterSE-RASePredRQ2}]{\scalebox{\scalingFactor}{
				\includegraphics[trim=0.5cm 0.5cm 0.4cm 1cm]{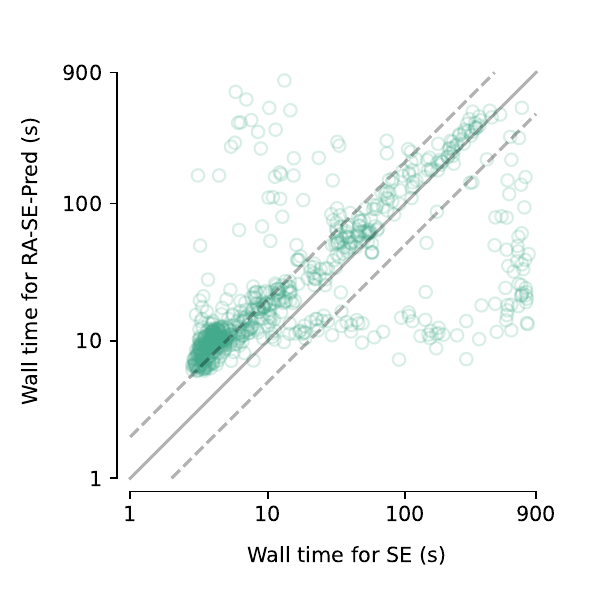}}
		}\hfil
		\subfloat[\Value and \RAValuePred\label{fig:scatterValue-RAValuePredRQ2}]{\scalebox{\scalingFactor}{
				\includegraphics[trim=0.5cm 0.5cm 0.4cm 1cm]{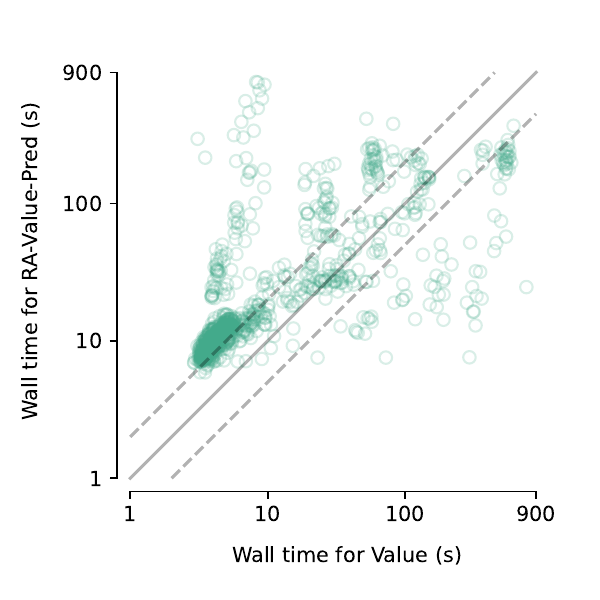}}
		}
		\hfil
		\subfloat[\BMC and \RABMCPred \label{fig:scatterBMC-RABMCPredRQ2}]{\scalebox{\scalingFactor}{
				\includegraphics[trim=0.5cm 0.5cm 0.4cm 1cm]{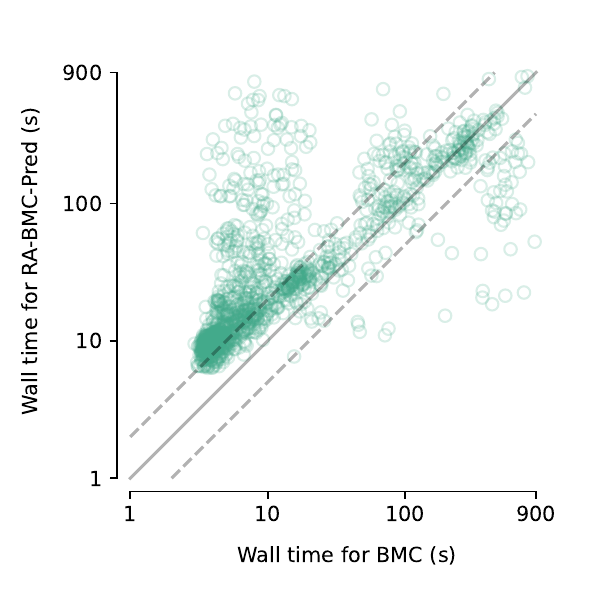}}
		}

		\hfil
		\subfloat[\Pred and \RASEPred\label{fig:scatterPred-RASePredRQ2}]{\scalebox{\scalingFactor}{
				\includegraphics[trim=0.5cm 0.5cm 0.4cm 1cm]{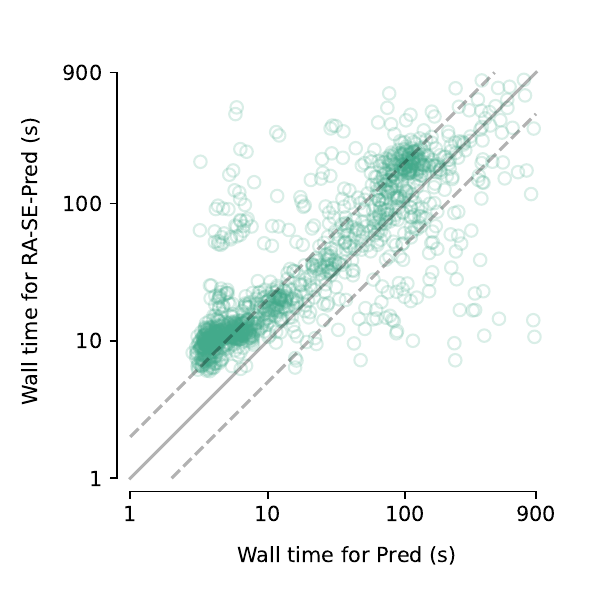}}
		}\hfil
		\subfloat[\Pred and \RAValuePred\label{fig:scatterPred-RAValuePredRQ2}]{\scalebox{\scalingFactor}{
				\includegraphics[trim=0.5cm 0.5cm 0.4cm 1cm]{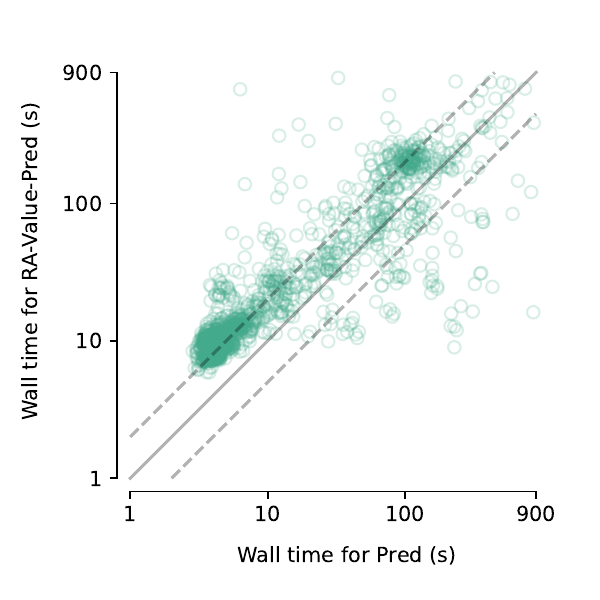}}
		}
		\hfil
		\subfloat[\Pred and \RABMCPred \label{fig:scatterPred-RABMCPredRQ2}]{\scalebox{\scalingFactor}{
				\includegraphics[trim=0.5cm 0.5cm 0.4cm 1cm]{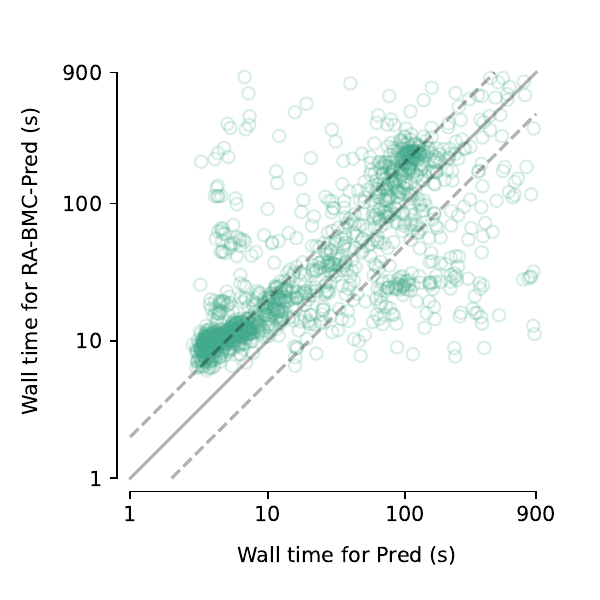}}
		}
		\vspace*{-3mm}
		\caption{Scatter plots comparing the wall time of the basic analyses and ranged program analysis using different analyses. \label{fig:scatter-rq2}}
	\end{minipage}
	\vspace*{-5mm}
\end{figure}

We make the same observations for the other two combinations of value analysis with predicate abstraction and for BMC with predicate abstraction:
	\RAValuePred has a higher effectiveness compared to value analysis, as the total number of solved tasks increases by {\moreSolvedVsFirstRAValuePred} (from  \num{\correctOverallValue} to \num{\correctOverallRAValuePred}).
	In contrast, predicate abstraction computes \num{\correctOverallPred} correct answers and thus \num{\moreSolvedVsOtherRAValuePred} more than \RAValuePred.
The reason for the lower effectiveness compared to predicate abstraction is again the fact, that the value analysis fails to solve its assigned range in most cases.
The ranged program analysis can solve \num{\additionalVsBothRAValuePred} tasks that neither predicate abstraction nor value analysis can solve.

BMC can solve more tasks than predicate abstraction on the benchmark (\num{\correctOverallBMC} resp. \num{\correctOverallPred})  and the combination \RABMCPred within ranged program analysis solves \num{\correctOverallRABMCPred} tasks.
Hence, \RABMCPred computes in total \num{\moreSolvedVsOtherRABMCPred} more correct answers than predicate abstraction, but \num{\moreSolvedVsFirstRABMCPred} fewer than BMC.
Nevertheless, \RABMCPred computes a correct solution for \num{\additionalVsBothRABMCPred} tasks, where both, predicate abstraction and BMC, do not succeed.

\begin{figure}[t]
	\begin{minipage}{\linewidth}
		\centering
		\captionsetup[subfigure]{justification=centering}
		\subfloat[\SymbExec and \Pred \label{fig:app:increaseSePredRQ2}]{\scalebox{0.5}{
				\includegraphics[trim=0.5cm 0.2cm 0.4cm 0.2cm]{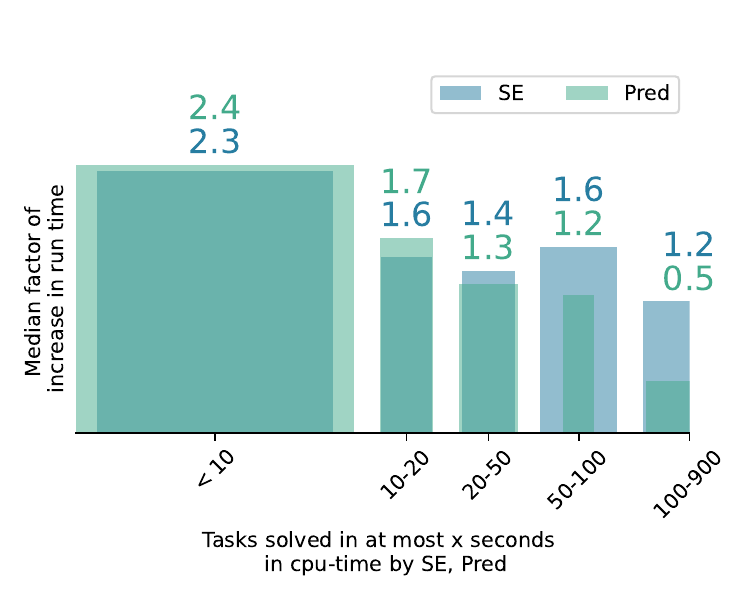}}
		} \hfil
		\subfloat[\Value and \Pred \label{fig:app:increaseValuePredRQ2}]{\scalebox{0.5}{
				\includegraphics[trim=0.5cm 0.2cm 0.4cm 0.2cm]{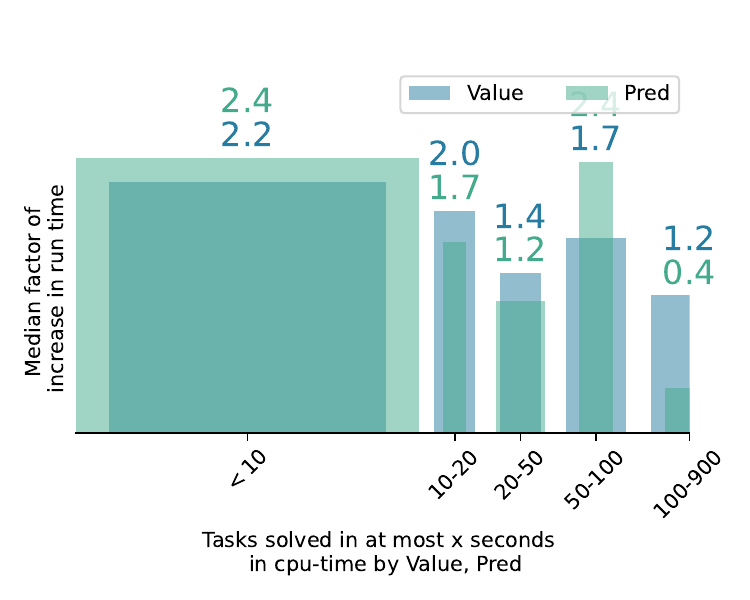}}
		}

		\subfloat[\BMC and \Pred \label{fig:app:increaseBMCPredRQ2}]{\scalebox{0.5}{
				\includegraphics[trim=0.5cm 0.2cm 0.4cm 0.2cm]{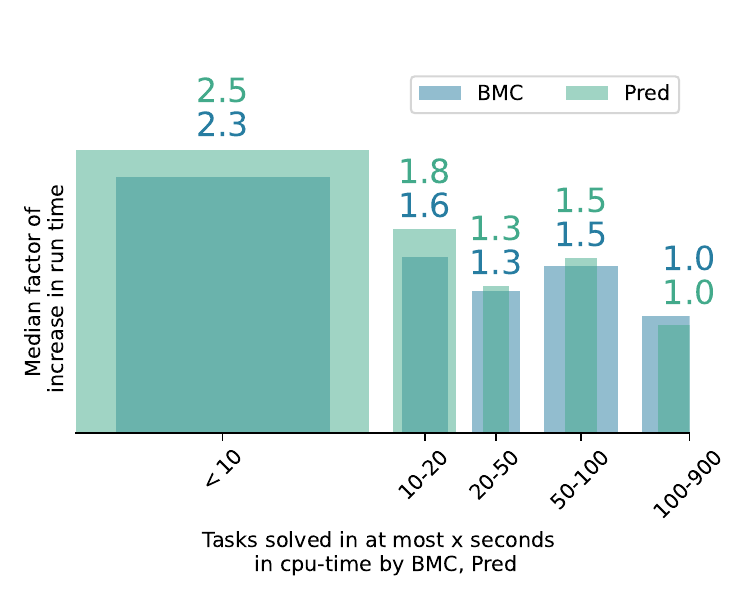}}
		}
	
		%
		
		\caption{Median wall~time increase of the default analyses compared to the ranged program analysis. The width of the bars corresponds to the number of tasks solved by the default analyses in that time interval. \label{fig:app:increaseRQ2}}
	\end{minipage}
	\vspace*{-5mm}
\end{figure}

\inlineheadingbf{Efficiency}
We depict the scatter plots comparing the wall~time of the basic analyses and the ranged program analysis in \cref{fig:scatter-rq2}.
For each ranged program analysis we present two scatter plots, one for a comparison with the first and one for the second basic analysis used.
We obtain similar results as in RQ~1:
The ranged program analysis needs approximately twice as long for simple tasks that can be solved within less than ten seconds and has a comparable overall run time for complex tasks.
In \cref{sec:RQ2}, we compare the efficiency of ranged program analysis using two different basic analyses.
In \cref{fig:app:increaseRQ2}, we present a more detailed comparison of the overall run time of the basic analyses with the ranged program analysis.

\inlineheadingbf{Summary}
All three configurations of ranged program analysis employing two different analyses in combination can solve tasks that are not solved by both basic analyses.
In addition, the combination has a higher effectiveness compared to the analysis that solves fewer tasks, but a lower one compared to the analysis solving more tasks.

%
%
%
%
%
%
%
%
%
%
\subsection{RQ 3: The Effect of Work Stealing}\label{sec:RQ3}
\inlineheadingbf{Evaluation Plan}
Next, we investigate the effect of work stealing.
	The idea of work stealing is motivated by the fact that the combination of different analyses within ranged program analyses has overall no positive effect on the effectiveness.
	We observe that these instances fail to solve tasks that one of the employed basic analyses can solve, as the other one does not complete the verification on the assigned range.
Therefore, we enable work stealing for each of the three configurations of ranged program analysis used for RQ~2, now denoted \WSSEPred, \WSValuePred, and \WSBMCPred.
We compare the results of the configurations using work stealing to the basic analyses and to ranged program analyses without work stealing and again compare effectiveness and efficiency.

\inlineheadingbf{Effectiveness}
To evaluate the effectiveness of ranged program analysis applying work stealing and two different analyses,
we have to compare the fourth block of \cref{tab:RQ1} against the first and third blocks.
The column \textit{par.only} in the fourth block now contains the tasks that neither of the two basic analyses can solve but the ranged program analysis using work stealing can, whereas the other columns are as explained in RQ~1.

\inlineheadingit{Work Stealing vs. Basic Analyses}
First and foremost, we observe that work stealing has a huge positive effect.
In comparison to the basic analyses, the ranged program analyses combining symbolic execution with predicate abstraction and BMC with predicate abstraction outperform both employed basic analyses when work stealing is enabled.
\WSSEPred can solve in total \num{\correctOverallWSSEPred} tasks, that are
\num{\moreSolvedVsFirstWSSEPred} tasks more than symbolic execution (solves \num{\correctOverallSE}) and \num{\moreSolvedVsOtherWSSEPred} tasks more than predicate abstraction (solves \num{\correctOverallPred}), an increase of  \num{\moreSolvedPercentageVsFirstWSSEPred}\% respectively  \num{\moreSolvedPercentageVsOtherWSSEPred}\%.
In addition, \WSBMCPred solves in total \num{\correctOverallWSBMCPred} tasks, that are \num{\moreSolvedVsFirstWSBMCPred} tasks more than BMC (\num{\correctOverallBMC}) and \num{\moreSolvedVsOtherWSBMCPred} more than predicate abstraction (\num{\correctOverallPred}),
increasing the effectiveness by \num{\moreSolvedPercentageVsFirstWSBMCPred}\% respectively \num{\moreSolvedPercentageVsOtherWSBMCPred}\%.

\begin{figure}[t]
	\begin{minipage}{\linewidth}
		\captionsetup[subfigure]{justification=centering}
		\subfloat[\RASEPred and \WSSEPred\label{fig:vennWS-SePredRQ3}]{\scalebox{\scalingFactorVenn}{
				\includegraphics[trim=0.5cm 0.5cm 0.4cm 1cm]{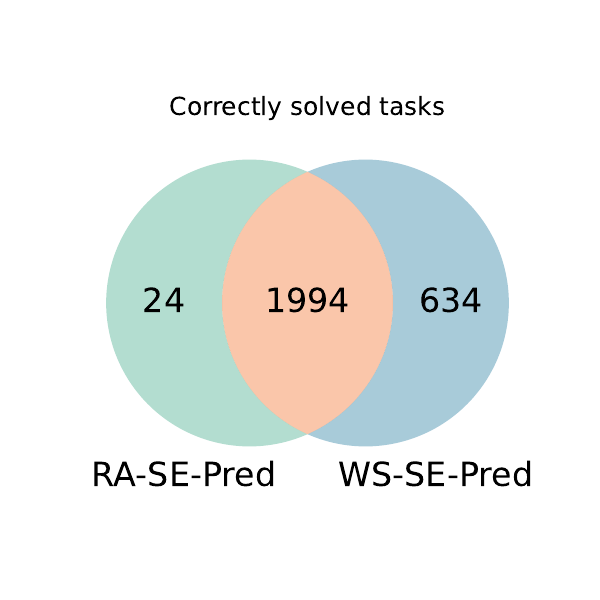}}
		}\hfil
		\subfloat[\RAValuePred and \WSValuePred\label{fig:vennWS-ValuePredRQ3}]{\scalebox{\scalingFactorVenn}{
				\includegraphics[trim=0.5cm 0.5cm 0.4cm 1cm]{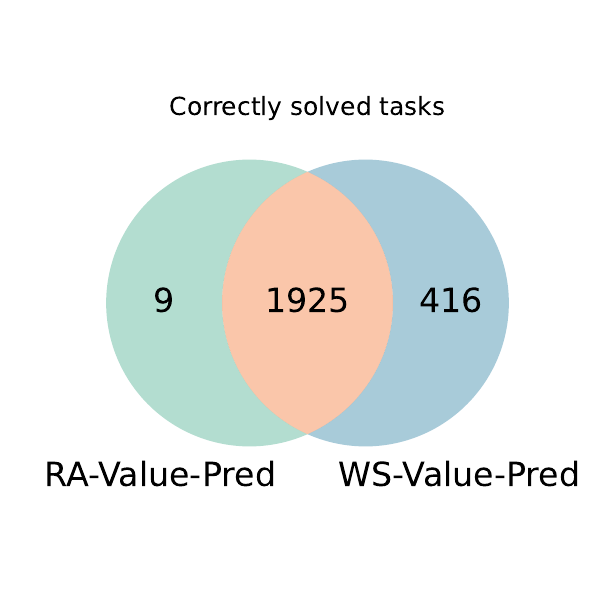}}
		}\hfil
		\subfloat[\RABMCPred and \WSBMCPred\label{fig:vennWS-BmcPredRQ3}]{\scalebox{\scalingFactorVenn}{
				\includegraphics[trim=0.5cm 0.5cm 0.4cm 1cm]{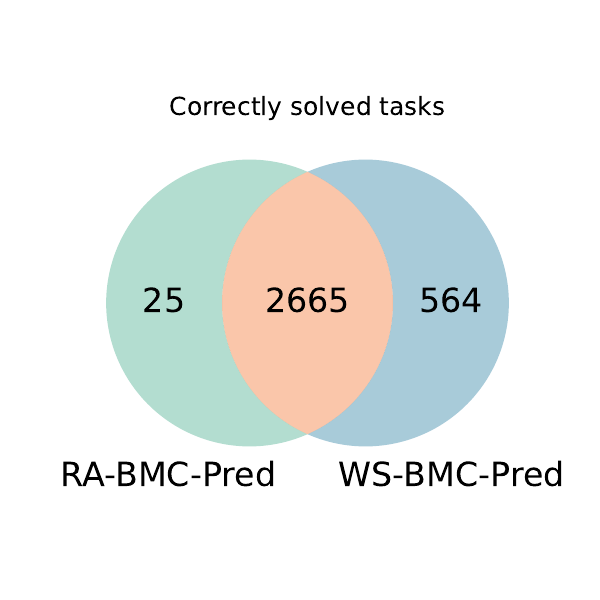}}
		}
		\vspace*{-3mm}
		\caption{Venn diagrams comparing the total number of correctly solved tasks using ranged program analysis with and without work stealing \label{fig:venn-rq3}}
	\end{minipage}
	\vspace*{-5mm}
\end{figure}

When combining value and predicate abstraction, work stealing also increases the effectiveness, although the positive effect is not as large as for the two other combinations:
	\WSValuePred solves \num{\correctOverallWSValuePred} tasks, \Value \num{\correctOverallValue} tasks, and \Pred \num{\correctOverallPred} tasks.
	Hence, \WSValuePred increases the number of solved tasks compared to \Value by \num{\moreSolvedVsFirstWSValuePred} and thus the effectiveness, but solves \num{\moreSolvedVsOtherWSValuePred}  tasks less compared to predicate analysis.
For the majority of these tasks, the splitter fails to generate a range and thus only the value analysis is executed on the full program.
In comparison to the value analysis, \WSValuePred is able to compute  \num{\moreSolvedVsFirstWSValuePred} more proofs and alarms than the default analysis.

\inlineheadingit{Work Stealing vs. No Work Stealing}
The positive effect of work stealing also becomes visible when comparing the total number of correct answers computed by ranged program analyses that do and do not employ it, as depicted in \cref{fig:venn-rq3}.
Each Venn diagram shown depicts on the left (in green) the number of tasks solved exclusively when using ranged program analysis without work stealing, on the right (in blue) the number of tasks solved exclusively when using work stealing and the intersection is labeled with the number of tasks solved by both.
Using work stealing enables the ranged program analyses to solve \num{\additionalVsFirstWSSEPredVsRA} additional tasks compared to \RASEPred,
\num{\additionalVsFirstWSValuePredVsRA} compared to \RAValuePred and \num{\additionalVsFirstWSBMCPredVsRA} compared to \RABMCPred.

Beneath the positive effect with respect to the number of correctly solved tasks, we can observe no negative effect in terms of incorrect answers.
When comparing the numbers of incorrect alarms computed when using work stealing (see third and fourth block of \cref{tab:RQ1}),
we note that no additional incorrect answers are computed.

\begin{figure}[t]
	\begin{minipage}{\linewidth}
		\captionsetup[subfigure]{justification=centering}
		\hfil
		\subfloat[\RASEPred and \WSSEPred \label{fig:scatterRASePred-WSSePredRQ3}]{\scalebox{\scalingFactor}{
				\includegraphics[trim=0.5cm 0.5cm 0.4cm 1cm]{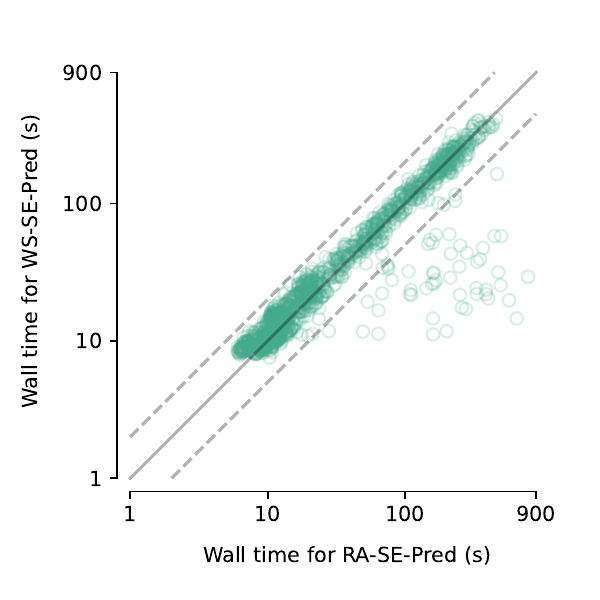}}
		}		\hfil
		\subfloat[\RAValuePred and \WSValuePred \label{fig:scatterRAValuePred-WSValuePredRQ3}]{\scalebox{\scalingFactor}{
				\includegraphics[trim=0.5cm 0.5cm 0.4cm 1cm]{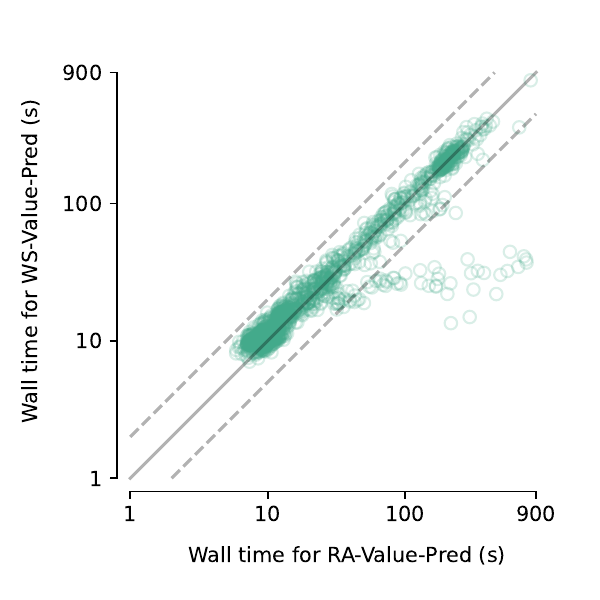}}
		}
		\hfil
		\subfloat[\RABMCPred and \WSBMCPred \label{fig:scatterRABMCPred-WSBMCPredRQ3}]{\scalebox{\scalingFactor}{
				\includegraphics[trim=0.5cm 0.5cm 0.4cm 1cm]{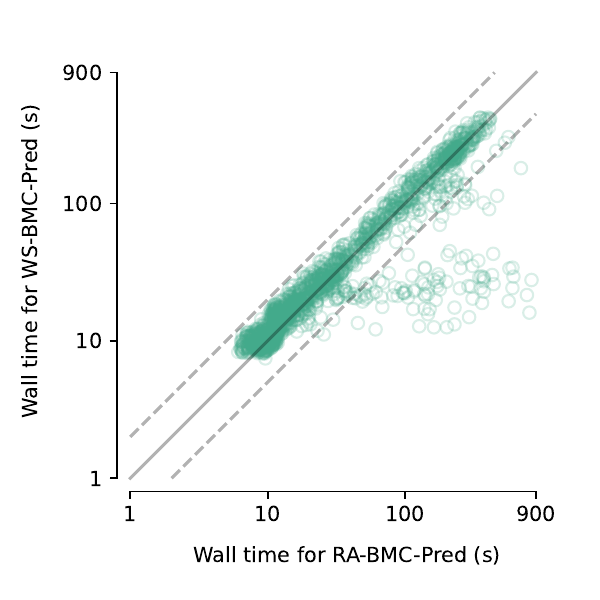}}
		}

		\vspace*{-3mm}
		\caption{Scatter plots comparing the wall time of the ranged program analyses using work stealing and no work stealing with different basic analyses. \label{fig:scatter-rq3}}
	\end{minipage}
	\vspace*{-5mm}
\end{figure}

\inlineheadingbf{Efficiency}
When comparing the efficiency of work stealing, we observe no negative effect compared to the normal ranged program analysis,
as depicted in \cref{fig:scatter-rq3}.
Therein, we observe that nearly all data points in the scatter plot are on or close to the diagonal, meaning that both analysis
take the same overall time to compute the answer.
In a few cases, work stealing can even decrease the overall time.\pagebreak

\inlineheadingbf{Summary}
The use of work stealing pays off.
With respect to effectiveness, we see that the combination of two best-performing basic analyses within ranged program analysis now outperforms the two basic analyses by \num{\moreSolvedPercentageVsFirstWSBMCPred}\% respectively \num{\moreSolvedPercentageVsOtherWSBMCPred}\%.
Moreover, work stealing has no negative effect in terms of efficiency compared to default ranged program analysis.

%
%
%
%
%
%
%
%
%
%

\subsection{RQ 4: Witness Joining in Ranged Program Analysis}\label{sec:RQ4}
\inlineheadingbf{Evaluation Plan}
Finally, we want to analyze whether the novel algorithm for witness joining generates correctness witnesses that can be validated.
For witness validation, we follow the schema used in the \svcomp and call a witness \emph{validated}, if there is at least one validator accepting the witness.
We employed the two best-performing correctness witness validators in \svcomp'23, namely \UA and \cpachecker.
We are again interested in effectiveness and efficiency.
As \cref{alg:joinWit} only joins correctness witnesses (violation witnesses do not need to be merged),
we selected all correct tasks for which the splitter \LBThree generates a range.
From the remaining tasks, we select these tasks where both value analysis and predicate abstraction generate a correctness witness that is validated, yielding in total \num{\numberTasksValidatedByBoth} tasks.

\inlineheadingbf{Effectiveness}
The ranged program analysis using value and predicate abstraction solve \num{\numberTasksSolvedRAOnReducedDataset} tasks, the remaining \num{6} tasks are not solved by \RAValuePred.
We validated all generated witnesses using \cpachecker and \UA, whereby \num{\numberTasksValidatedRAOnReducedDataset} (\num{\percentagValidatedWitnessesRAOnReducedDataset}\%) of the joined correctness witnesses where validated.
The remaining \num{\numberTasksNotValidatedRAOnReducedDataset} tasks are not validated, as the validator reaches the memory limitations during the validation.
A manual inspection of the witnesses shows that these witnesses are also valid.

When comparing the effectiveness of \RAValuePred to predicate abstraction and value analysis (cf. \cref{tab:RQ1}),
we notice that there are \num{\additionalVsBothRAValuePred} tasks only solved by the ranged program analysis, whereas \num{6} are additional proofs.
From these \num{6} additional proofs \num{5} are validated.
One additional proof is not validated by the validators, as they reach the memory limitations during validation.\footnote{Due to the size of the generated witness a manual inspection is infeasible.}

\begin{figure}[t]
	\begin{minipage}{\linewidth}
		\captionsetup[subfigure]{justification=centering}
		\hfil
		\subfloat[The scatter plot\label{fig:scatterWitnessJoin}]{\scalebox{\scalingFactor}{
				\includegraphics[trim=0.5cm 0.5cm 0.4cm 1cm]{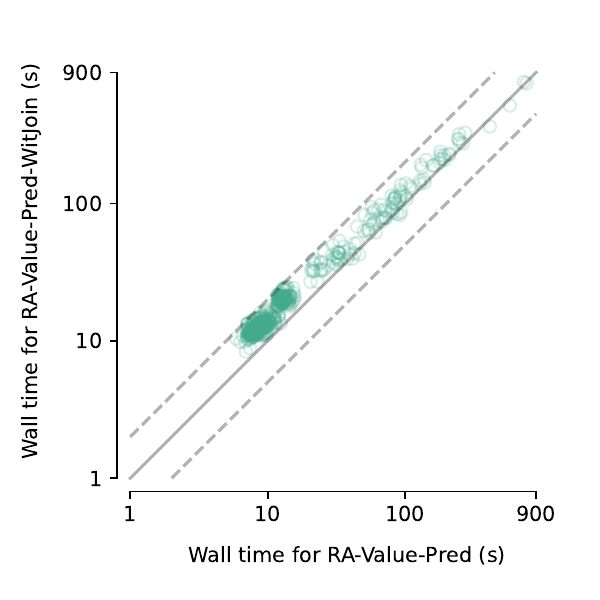}}
		}		\hfil
		\subfloat[Median wall~time increase of \RAValuePred with witness joining.
			\label{fig:comparisonWitnessJoin}]{\scalebox{\scalingFactor}{
				\includegraphics[trim=-2cm 0.4cm -2cm 1.8cm]{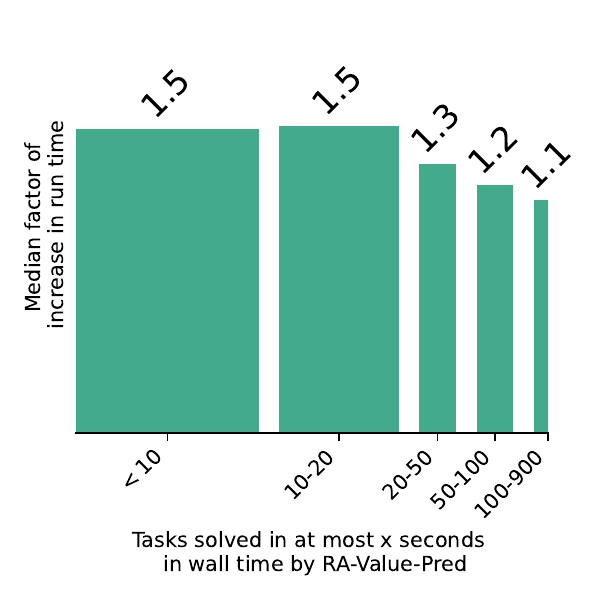}}
		}
		\hfil
		\vspace*{-3mm}
		\caption{Comparison of \RAValuePred with and without witness joining. \label{fig:scatter-rq4}}
	\end{minipage}
	\vspace*{-5mm}
\end{figure}

\inlineheadingbf{Efficiency}
Beneath the quality of the joined verification witnesses, we are also interested in the additional overhead caused by the application of the witness join.
We compare in \cref{fig:scatterWitnessJoin} the overall time taken by \RAValuePred, where we depict on the x-axis the time of the analysis when joining of the witnesses is disabled and on the y-axis with enabled witness joining.
As expected, we observe that computing the joined correctness witness increases the overall time.
For a more detailed analysis, we group the tasks by the time taken by the analysis without witness join and compute for each group the median increase in the run time, visualized in \cref{fig:comparisonWitnessJoin}.
We notice that the additional overhead caused by joining the witnesses decreases for tasks that need more time to solve.
For tasks that take more than \num{100} seconds to solve, the additional time needed to compute a solution is only around \num{10}\% of the overall computation time.

\inlineheadingbf{Summary}
The algorithm for witness joining proposed in \cref{alg:joinWit} works, as it generates witnesses that are in more than
\num{99}\% of the cases successfully validated.

\subsection{Threats to Validity}
\inlineheadingbf{Internal Validity}
It is unlikely that the implementation of the ranged program analysis suffers from bugs, as we obtain comparable results for the ranged program analysis in~\cite{RangedAnalysisSEFM}, where the ranged program analysis is
realized using instrumentation instead of the CPA-based realization that we use in this paper.
The additional incorrect answers computed for the \RASE are most likely not caused by an error in the ranged program analysis,
as they are also observed in \cite{RangedAnalysisSEFM}.
Moreover, we randomly selected some of the additionally raised incorrect alarms and verified, that the basic analyses also report the alarms, if the search space is limited to the path that was incorrectly reported by the ranged program analysis.

The data collected may deviate when conducting a reproduction study using the same experimental setup.
To account for small, expected measurement errors, we restrict the presentation of our data to two significant digits.

\inlineheadingbf{External Validity}
We have conducted the experiments using the \svbenchmarks, one of the largest available datasets for C program verification.
Although it is widely used, especially in the competition \svcomp, our findings may not completely carry over to other real-world C programs.

In our experiments, we mainly focused on tasks from the \texttt{ReachSafety} category of the \svbenchmarks. 
To assess the generalizability to other tasks, we additionally evaluated the best-performing configuration \RABMCPred on the software-systems category from \svcomp, which comprises of additional \num{2985} tasks. 
The tasks in the software-systems category are designed to be more representative of real-world use cases and include realistic implementations from the AWS C commons library or the Linux kernel. 
We observe that the employed configuration of BMC scales poorly to the software-systems tasks, especially since it performs worse than predicate abstraction in detecting property violations\footnote{The results are contained in our artifact~\cite{artifact}.}. 
As a result, the combination of BMC and predicate abstraction within \RABMCPred is deemed to be less effective compared to predicate abstraction. 
We still observe that work stealing has a positive effect when employed in ranged program analysis, as it increases the number of correctly solved tasks by around \num{44}\% compared to \RABMCPred.

We did not redo the experiments from~\cite{RangedAnalysisFASE}, where we compared the different splitting strategies for the configurations of ranged program analysis.
In our prior experiments, \LBThree was the splitter the ranged program analyses achieved the best performance with.
It might be the case that certain combinations of ranged program analyses perform even better when a different splitting strategy is used.
In that case, the findings from our evaluation and conclusions drawn still remain valid.

The concept of ranged program analysis presented in this work is conceptually applicable to arbitrary program analyses.
The range reduction presented in \cref{sec:range-reduction} can only be applied to analyses that are implemented in the framework of configurable program analysis.
To overcome this limitation, we have presented a more generic approach based on program instrumentation in~\cite{RangedAnalysisSEFM}, that allows to use of arbitrary off-the-shelf program analyses.
The instrumentation-based construction of ranged analyses can also be used in combination with work stealing.
Combining correctness witnesses, work stealing, and instrumentation is planned as future work.

\inlineheadingbf{Data Availability Statement}

All experimental data and our open-source implementation are publicly available in our supplementary artifact~\cite{artifact} ar\-chived at \zenodo.
Therein, we also included a virtual machine for reproducing the experiments. 

\section{Related Work}
Ranged program analysis runs different analyses on separate partitions of the search space, but accompanies this with work stealing.
After the analyses are finished, it joins their result.
In the following, we discuss related work for all those aspects in individual subsections.

\subsection{Analysis Combinations}
Different strategies exist to combine different analyses or analysis tools.
For example, one may use an analysis or tool as a component in another approach~\cite{SymbolicSearchTest,EvoSE,CoqPlusGappa,SUSHI,VeriFuzz,DBLP:conf/cav/ChristakisEHHKL21,EvosuiteCombi,SynergiSE,BoostSearchTest}.
Furthermore, one may follow the idea of the strategy design pattern~\cite{StrategyPattern} and perform algorithm selection~\cite{AlgorithmSelection} to choose an appropriate analysis from a set of analyses~\cite{StrategySelection,PredictingRankings,PortfolioVerificationCAV15,DBLP:conf/icst/GargantiniV15,DBLP:conf/icse/JiaCHP15,PredictingJournal,MUX}.
Other strategies control the execution order of the combination.
For instance, the different analyses can be executed sequentially.
Sequential portfolio approaches~\cite{CPAseq,Ultimate-Portfolio} define an order of the analyses and continue with the next analysis if the previous analysis did not succeed, e.g., due to time out or failure.
Testification is the process of looking for evidence to back up the claim that a property is violated or fulfilled.
Thereby, testification approaches~\cite{SANTE,DynaBoost,CheckNCrash,Orion,DyTa,ResidualInvestigation,AIPlusBMC} should make the analysis result more precise.
Hence, they typically combine a verification and a validation approach and prioritize or only report confirmed proofs and alarms. 
To increase the cooperation among sequential combinations even further, some sequential combinations~\cite{ConditionalTesting,FuSeBMC21,DBLP:conf/pts/HusterSRKR17,ConditionalModelChecking,DBLP:conf/icse/BeyerJLW18,ChristakisMW16,DBLP:conf/fm/ChristakisMW12,CzechConditionalModelChecking,ProgramTrimming,ALPACA,ProgramPartitioning} restrict the subsequent analyses to the yet unexplored state space.
An extension of sequential combinations is to iteratively run sequential combinations resulting in so called interleaved combinations~\cite{UnderOverApproximations,Veritesting,DASH,CoVeriTest,AbstractionConcolicTesting,ConditionalTesting,LLSPAT,MayMustAnalysis,Synergy,Opal,EvaconCombi,HybridConcolicTesting,Badger,Driller,YorshSagivBall}. 
Typically, interleaved combinations may exchange some kind of information. 
An alternative to sequential or interleaved combination is the parallel combination, which we use in our ranged program analysis.
The simplest parallel combination is a parallel portfolio~\cite{DBLP:conf/sefm/TschannenFNM11,ConditionalTesting,SwarmTest,PredatorTool,SwarmVerification,SCAR} that runs different analysis independently in parallel.
More refined parallel combinations work together.
They may exchange information encountered during analysis~\cite{kInduction,CPA,CousotPOPL79,CousotCombinations,DART,SAGE,CoVEGI,HiDiff}.  
Also, they may distribute the search space among different instances of the same analysis~\cite{ParallelKorat,DBLP:conf/icse/WeiJGDTBR23,SynergiSE,Sym-Range,DBLP:conf/icfem/SinghK20,DBLP:journals/corr/abs-2106-02179,Sym-SSP,SynergiSE-journal,RangedModelChecking,DBLP:conf/ppopp/InversoT20,CSEQ-SWARM,StructurallyDefinedCMC}.
Our ranged program analyses and instrumentation-based ranged program analysis~\cite{RangedAnalysisSEFM}, which encodes the range constraints into the program instead of running a range reduction in parallel to the analysis, go beyond the latter and allow to distribute the search space among different analyses.

\subsection{Search Space Partitioning}
To let analyses explore different parts of the search space, the search space needs to be partitioned.
Partitioning can be done dynamically based on the already performed search or statically before the exploration.
A common dynamic partitioning is to consider the already performed analysis and divide the search space into an already covered and a yet uncovered part.
Thereby, the search space may be described based on test goals~\cite{ConditionalTesting,FuSeBMC21}, open proof obligations~\cite{DBLP:conf/pts/HusterSRKR17}, a previous analysis result~\cite{DBLP:conf/sas/YinCL0C19}, program paths~\cite{ConditionalModelChecking,CoVeriTest,DBLP:conf/icse/BeyerJLW18,ChristakisMW16,DBLP:conf/fm/ChristakisMW12,CzechConditionalModelChecking,AbstractionConcolicTesting,ProgramTrimming,ALPACA,ProgramPartitioning}, or the already explored symbolic execution tree~\cite{Cloud9-EuroSys,Cloud9,SynergiSE,Sym-Range,SECloud,DBLP:conf/icse/WeiJGDTBR23}.
Static partitioning may for instance divide the analysis into separate subtasks~\cite{CoDiDroid,DBLP:journals/sigsoft/YangDW15,ConditionalTesting} or use a static state partitioning function~\cite{BAM-parallel,MurPhi,ParallelSPIN,DBLP:conf/spin/GaravelMS01,DBLP:conf/kbse/BarnatBC03} to divide the states into different sets.
Another partitioning strategy is to split the program paths.
Conditional static analysis~\cite{StructurallyDefinedCMC} uses the order of executed program branches to characterize the different sets of program paths.
Similarly, some ranged symbolic execution approaches~\cite{DBLP:conf/icfem/SinghK20,DBLP:journals/corr/abs-2106-02179}~\cite{RangedModelChecking} characterize the sets of paths via path prefixes.
In contrast, one ranged symbolic execution approach~\cite{Sym-SSP} defines input constraints to define the sets of paths.
Korat~\cite{ParallelKorat} defines input ranges while our approach, some ranged symbolic execution approaches~\cite{SynergiSE,Sym-Range,SynergiSE-journal}, and ranged model checking specify path ranges.
    In contrast to path ranges, the set of paths described by an input range is not guaranteed to be ordered.
    Thus, the technique for range reduction presented may not work.
Ranged symbolic execution creates path ranges from an initial shallow symbolic execution~\cite{DBLP:conf/icfem/SinghK20,DBLP:journals/corr/abs-2106-02179,Sym-SSP} or tests~\cite{SynergiSE,Sym-Range,SynergiSE-journal}.
Our ranged program analysis uses splitting strategies, e.g., based on randomization or loop unrollings. 
Instrumentation-based ranged program analysis~\cite{RangedAnalysisSEFM} reuses our splitting strategy based on loop unrollings.

\subsection{Load Balancing} 
To be efficient, the workload of each parallel worker should be roughly the same.
There exist approaches that try to achieve an even workload distribution of the static partitioning~\cite{ParallelKorat,MurPhi,ParallelSPIN,DBLP:conf/spin/GaravelMS01,DBLP:conf/kbse/BarnatBC03}.
Nevertheless, often workload needs to be redistributed dynamically.
Like us, several approaches~\cite{DBLP:conf/icfem/SinghK20,Sym-Range,DBLP:journals/corr/abs-2106-02179,SynergiSE-journal} provide an idle worker with new work.
Other approaches~\cite{Cloud9-EuroSys,Cloud9,DBLP:journals/entcs/KumarM05} redistribute work from overloaded workers to workers with lower workload.
While our approach simply lets a worker analyze the state space not analyzed so far after it becomes idle, existing approaches typically take over some of the work of another worker.
To this end, several approaches offload the analysis of one subtree~\cite{DBLP:conf/icfem/SinghK20,Sym-Range,DBLP:journals/corr/abs-2106-02179}, several subtrees~\cite{Cloud9-EuroSys,Cloud9} or several states~\cite{DBLP:journals/entcs/KumarM05} that need to be analyzed.
In contrast, dynamic range refinement~\cite{SynergiSE-journal} first computes the yet unexplored range and splits it into two ranges, one being offloaded and the other being the new range to be analyzed.

\subsection{Result Aggregation}
The aggregation of verdicts is rather common in analysis combinations.
Analysis combinations like~\cite{PredatorTool,Veritesting,CSEQ-SWARM,DBLP:conf/ppopp/InversoT20} that underapproximate the program behavior typically only report if a violation is found by any of the analyses and possibly also the corresponding counterexample.
Analysis combinations like~\cite{UnderOverApproximations,DASH,Synergy} that combine under- and overapproximation analyzers each exploring the complete program report a counterexample found by any of the underapproximating analyzers and proofs found by the overapproximating verifiers.
Like our ranged program analysis, analysis combinations~\cite{ConditionalModelChecking,DBLP:conf/icse/BeyerJLW18,ChristakisMW16,DBLP:conf/fm/ChristakisMW12,CzechConditionalModelChecking,AbstractionConcolicTesting,ProgramTrimming,ALPACA,RangedAnalysisSEFM} that distribute the search space among different analyses or analysis instances often announce violations directly but report proofs only after all analysis runs are successfully finished. 
In contrast, SCAR~\cite{SCAR} outputs a quantitative value to describe the (in)correctness for each of its results.

While the aggregation of verdicts is common, aggregating the evidence for results of distributed search space exploration is discussed rarely.
For ranged program analysis, we introduced the aggregation of witnesses generated for subsets of the state space.
The introduced method is in general independent of ranged program analysis and can be used in other contexts.
Jakobs~\cite{ARGJoin} describes an approach to join partial analysis results in form of partial abstract reachability graphs, a description of the state space explored by an abstraction-based analyzer.
Garavel~\cite{DBLP:conf/spin/GaravelMS01} describe how to combine partial labeled transition systems obtained by distributive explicit state space exploration.

\section{Conclusion}
This paper introduces ranged program analysis, a technique to analyze a program by running different (configurable program) analyses in parallel on different path ranges. 
Our approach generalizes previous techniques by allowing for the combination of arbitrary analyses, 
including the joining of verification results at the end. 

Our experimental results show that the efficiency of ranged program analysis is comparable to the efficiency of the individual analyses.
Moreover, the usage of ranged program analysis allows for solving additional tasks, not solvable by individual analyses alone.
The experiments further confirmed the soundness of our algorithm for witness joining, by showing that standard witness validators are able to successfully validate our witnesses.

Our experiments suggest that combining a technique whose strength is finding property violations with another one good at computing proofs within ranged program analysis and employing work stealing can form a successful combination that can increase the overall effectiveness. 
Validating whether novel combinations using program instrumentation for off-the-shelf verifiers also benefit from using work stealing is one of the challenges that we aim to address next.
For instance, using complete tools like \esbmc~\cite{DBLP:conf/kbse/GadelhaMMC0N18}, \mopsa~\cite{DBLP:conf/vmcai/MilaneseM24}, or \ultimate~\cite{DBLP:conf/cav/HeizmannHP13} instead of individual analyses in combination seems promising.

Beneath work stealing the generation of the program ranges can influence the performance of ranged program analysis.
Currently, we employ \LBThree, a splitter that generates the ranges using a fixed heuristic.
Finding a more sophisticated technique that may also take the given task into account is a promising line of work, e.g. by using constant values extracted from the program or estimating the number of loop iterations for computing more balanced ranges.
Recently, the idea of dynamic, demand-driven splitting is realized in the tool \tool{Bubaak-SpLit}~\cite{DBLP:conf/tacas/ChalupaR24}.
Instead of computing the ranges up-front, \tool{Bubaak-SpLit} generates ranges in case configurations of \tool{Bubaak} fail to solve tasks within certain limits.
This idea opens another promising line of work for ranged program analysis.

    Lastly, the information computed by one ranged analysis may also be helpful for the other analysis running in parallel.
    Hence, we aim to establish information exchange among ranged analyses, to enable cooperation within ranged program analysis.
    Beneath that, we aim to evaluate if generating more than one counterexample within ranged program analysis is beneficial.

\inlineheadingbf{Data Availability Statement}
All experimental data and our open-source implementation are publicly
available in our supplementary artifact~\cite{artifact}.
Therein, we also included a virtual machine for reproducing the experiments.

\bibliographystyle{elsarticle-num}
\bibliography{literature}

\end{document}